\documentclass[11pt]{article}
\begin{document}
\def\a{\alpha}\def\b{\beta}\def\g{\gamma}\def\d{\delta}\def\e{\epsilon }
\def\k{\kappa}\def\l{\lambda}\def\L{\Lambda}\def\s{\sigma}\def\S{\Sigma}
\def\Th{\Theta}\def\th{\theta}\def\om{\omega}\def\Om{\Omega}\def\G{\Gamma}
\def\y{\vartheta}\def\m{\mu}\def\n{\nu}
\renewcommand\baselinestretch{1.2}
\newcommand{\nn}{\nonumber\\}\newcommand{\p}[1]{(\ref{#1})}
\renewcommand{\thefootnote}{\fnsymbol{footnote}}
\thispagestyle{empty}
\begin{flushright}
March 8, 2003,  hep-th/0303075. 
\\ JHEP 0308 (2003) 032 
\end{flushright}

\vspace{3cm}

\begin{center}

{\Large\bf SUPERFIELD T--DUALITY  RULES}

\vspace{0.2cm}
{\bf Igor A. Bandos $^\dagger{}^\ast$ and Bernard Julia $^\ddagger$} 

\vspace{0.5cm}
{\it $^\dagger$Departamento de F\'{\i}sica Te\'orica and IFIC 
(CSIC-UVEG), 
 46100-Burjassot (Valencia), Spain}

\vspace{0.5cm}
{\it $^\ddagger$ Laboratoire de Physique Th\'eorique de l'Ecole Normale 
Sup\'erieure, 75231 Paris Cedex 05, France: Bernard.Julia@ens.fr}

\vspace{0.5cm}
{\it $^\ast$Institute for Theoretical Physics, NSC KIPT, 
UA61108,
Kharkov, Ukraine}

\vspace{3.0cm}

{\bf Abstract}
\end{center}

A geometric treatment of T--duality as an operation which 
acts on differential forms in superspace 
allows us to derive 
the complete set of T-duality transformation rules which relate
the superfield potentials of $D=10$ type IIB supergravity with those of type 
IIA supergravity including Ramond--Ramond superfield potentials
($C^{(2n)}_{M_1\ldots M_{2n}}(Z)$ resp. $
\hat{C}^{(2n+1)}_{M_1\ldots M_{2n+1}}(\hat{Z})$) and 
fermionic supervielbeins ($E_M^{\alpha 1, 2}$ resp. 
$\hat{E}_M^{\alpha 1}, \hat{E}_{M\alpha}^{\; 2}$). We show that these 
rules are consistent with the superspace supergravity constraints.

\bigskip

\setcounter{page}0 

\newpage
\renewcommand{\thefootnote}{\arabic{footnote}}
\setcounter{footnote}0
\renewcommand{\theequation}{\arabic{section}.\arabic{equation}} 
\setcounter{equation}0

\section{Introduction}

T--duality is a perturbative symmetry of string theory which relates, 
for instance, the  
type IIA and type IIB superstring models (see, {\it e.g.}, \cite{Polch0}). 
The study of the bosonic string action in a 
background 
admitting an isometry \cite{buscher} provided an elegant representation 
of T--duality as a map between two spacetime
field theories. Such a field theoretical representation of 
T-duality was studied for the  bosonic limit of supergravity 
\cite{duff1,bho,bdr,ght,hull,lozano,zumino,myers,simon}. 
Progress on its supersymmetric generalization has been achieved only 
recently \cite{clps,hassan,kulik} (see \cite{siegel,bakas} for earlier
study). 
In \cite{clps} the study of the T-duality map of the 
component-field expansion of the Green-Schwarz action for the type IIA 
superstring up to quadratic order in the fermionic coordinate functions 
$\hat{\theta}(\tau,\sigma)$ 
was used to derive the type IIB superstring action with 
the same accuracy and, then, the model for 'massive type IIA superstring' 
(superstring in the Roman's massive type IIA  supergravity background). 
The T-duality rules for gravitini have been found in \cite{hassan}. 
Finally, in \cite{kulik} the  T--duality rules 
for NS--NS superfields [$E_M^a(Z)$, $B_{MN}(Z)$ and 
$\hat{E}_M^a(\hat{Z})$, $\hat{B}_{MN}(\hat{Z})$] 
and fermionic supervielbeins ($E_M^{\alpha 1, 2}$ and 
$\hat{E}_M^{\alpha 1}, \hat{E}_{M\alpha}^{\; 2}$) 
were found by studying the relation between complete 
type IIA and type IIB superstring actions and their $\kappa$--symmetries
However, this approach did not allow to find the T--duality rules for 
Ramond--Ramond (RR) {\sl superfield potentials} 
($C^{(2n)}_{M_1\ldots M_{2n}}(Z)$ and $
\hat{C}^{(2n+1)}_{M_1\ldots M_{2n+1}}(\hat{Z})$)
and required significant efforts to extract the transformation rules for the 
 components of the RR field strengths from Bianchi identities. 

One of the messages of this paper is that the {\it complete} set of 
superspace T-duality rules 
(including the rules for the RR superfield potentials)
can be obtained from the relation between 
the complete $\kappa$--symmetric actions 
for Dirichlet superbranes in type IIA and type IIB 
supergravity backgrounds 
and subsequent study of the
exchange between the type IIA and IIB superspace supergravity constraints.  
Namely, in the first stage, the comparison of the 
type IIA super--D$(p+1)$--brane and type IIB 
super--Dp--brane actions \cite{c0}, 
which are known to be related by T--duality \cite{Polch0,Polch}, 
provides  the T--duality transformation rules for all 
the {\it bosonic} superforms 
of type IIB resp. IIA supergravity:   
bosonic supervielbeins ($E^a(Z)$ resp. $\hat{E}^a(\hat{Z})$), 
NS-NS superforms ($B_2(Z)$ resp. $\hat{B}_2(\hat{Z})$)
and all RR superforms ($C_{2n}(Z)$ resp. $\hat{C}_{2n\pm 1}(\hat{Z})$).   
Then, in a second stage, 
substituting these rules into the 
superspace torsion constraints and the constraints 
on NS--NS field strengths of type IIA and type IIB supergravities 
\cite{HW84,CGO87,c0}, one can derive the T--duality rules for the remaining 
(fermionic) supervielbein forms.  

It turns out that the 
T-duality transformation rules for the bosonic superforms, which 
can be obtained from the comparison of  
the super--Dp--brane actions ({\it i.e.} by the 
superfield generalization of the {\sl method} of Ref. \cite{simon}), 
can be reproduced as well by a straightforward 
superfield (superform) generalization of the 
{\sl final results} of 
Ref. \cite{simon}
\footnote{Such a simple possibility to reproduce the 
superfield results from the component ones can be regarded as a reflection 
of the existence of the `rheonomic' 
(group manifold) approach to supergravity \cite{rheo} (see \cite{bsv} for its 
superbrane generalization) which allows to lift the component equations 
(written 
in terms of differential forms on spacetime) to the superspace equations 
for superforms. This is also natural in a view of 
recent observation \cite{BdAIL2} that superfield description of the 
dynamical supergravity--superbrane interacting system (still hypothetical  
for $D=10,11$) is gauge equivalent to a more simple dynamical system 
described by the sum of the standard (component) supergravity action 
and the action for pure bosonic brane 
(the pure bosonic limit of the original superbrane action).}. 
In this paper we shall use such a shortcut, it will allow us, in particular, 
to simplify notations. 
By substituting then the NS--NS T--duality rules thus obtained 
into the superspace torsion constraints and into the constraints 
on NS--NS field strengths
\cite{HW84,CGO87,c0}, we derive the T--duality rules for fermionic 
supervielbein forms in Einstein frame\footnote{Note that, although the authors 
of \cite{kulik} worked in the so--called string frame, 
where the superstring action does not include the dilaton superfield,  
one can verify 
that the requirement of superstring $\kappa$--symmetry in the 
presence of standard type II supergravity constraints \cite{c0} 
(see Eqs. (\ref{TaIIA=})--(\ref{H3IIB}) below, or, equivalently, 
Ref. \cite{kulik}: Eqs. (A.8), (A.9), (A.19)) would result in the 
trivialization of supergravity background {\sl if} one were to drop the 
dilaton factor from the superstring action. 
This indicates that the standard superfield supergravity constraints 
have been formulated in the so--called 
Einstein frame rather than in the string frame.
Hence, the superfield T--duality rules in the Einstein frame shall be more  
accessible and more useful for applications.}. 
Finally, we describe the verification  
of the consistency of the complete set of 
T--duality rules thus obtained with the superspace constraints for RR 
superform field strengths \cite{c0}. The T-duality transformation rules are collected 
in an appendix.

\section{Basic notions and notations}

The tangent space metric is mostly minus, 
$\eta_{\hat{a}\hat{b}}= diag (+1,-1,\ldots , -1) = \eta_{ab}$. 
The hat symbol 
$\;\hat{}\;$ is used to distinguish  superfields, coordinates and indices of   
the type IIA supergravity; 
the superfields and coordinates  
of type IIB superspace are denoted by the same symbols, but without hat. 
Bosonic supervielbein forms of type IIA and type IIB supergravity are 
denoted by 
\begin{equation}
\label{bEIIA}
type ~IIA: \qquad 
\hat{E}^{ {\hat{a}}}= d\hat{Z}^{ {\hat{M}}}
\hat{E}_{ {\hat{M}}}^{ {\hat{a}}}(\hat{Z}),
\end{equation}
\begin{equation}\label{bEIIB}
type ~IIB: \qquad 
E^{ {a}}= dZ^{ {M}} E_{ {M}}^{ {a}}(Z)\; , 
\end{equation}
NS--NS gauge {\sl superforms} by   
\begin{eqnarray}
\label{0B2IIA}
& type ~IIA: \; & \hat{B}_2
={1\over 2}
d\hat{Z}^{\hat{ {N}}} \wedge
d\hat{Z}^{\hat{ {M}}}
\hat{B}_{\hat{ {M}}\hat{ {N}}}(\hat{Z})
={1\over 2}
\hat{E}^{\hat{ {B}}} \wedge
\hat{E}^{\hat{ {A}}}
\hat{B}_{\hat{ {A}}\hat{ {B}}}(\hat{Z})\; , 
\\ \label{0B2IIB}
& type ~IIB: \; & B_2
={1\over 2}
dZ^{ {N}} \wedge
dZ^{ {M}}B_{ {M} {N}}(Z)
={1\over 2}
E^{ {B}} \wedge
E^{ {A}}B_{ {A} {B}}(Z)\; , 
\end{eqnarray} 
and the fermionic supervielbein forms by   
\begin{eqnarray}
\label{fEIIA}
type ~IIA: &  
\hat{E}^{ {\hat{\a}}}=
(\hat{E}^{ {\a}1}, \hat{E}^2_{ {\a}})\; , 
  \quad &  \hat{E}^{ {\a}1}=  d\hat{Z}^{ {\hat{M}}}
\hat{E}_{ {\hat{M}}}^{ {\a}1}(\hat{Z})\; , 
\quad \hat{E}^2_{ {\a}}
=  d\hat{Z}^{ {\hat{M}}}
\hat{E}_{ {\hat{M}}}{}^2_{ {\a}}(\hat{Z}) \; , \quad 
\\ 
\label{fEIIB}
 type ~IIB:  &  
E^{ {\breve{\a}}}= (E^{ {\a}1}, E^{ {\a}2}) \; , 
\quad & 
E^{ {\a}1}= dZ^{ {M}} E_{ {M}}^{ {\a}1}
(Z)\, ,  \quad 
E^{ {\a}2}= dZ^{ {M}} E_{ {M}}^{ {\a}2}(Z)
\, .  \quad 
\end{eqnarray}
Here $\alpha = 1, \ldots , 16$ is $D=10$ Majorana--Weyl spinor index. 
Upper and lower indices correspond to opposite  chiralities. 
Ten dimensional $16\times 16$ sigma matrices, 
$\sigma_{\alpha\beta}^{a}$, $\tilde{\sigma}^{a\alpha\beta}$ 
are real, symmetric and 
 satisfy $(\sigma^a \tilde{\sigma}^b + 
\sigma^b \tilde{\sigma}^a)_\alpha{}^\beta = 2\eta^{ab}\delta_\alpha{}^\beta$; 
$\sigma^{ab}= \sigma^{[a} \tilde{\sigma}^{b]}= 1/2 
(\sigma^a \tilde{\sigma}^b -
\sigma^b \tilde{\sigma}^a)$, 
$\tilde{\sigma}^{ab}= \tilde{\sigma}^{[a} {\sigma}^{b]}$, 
$\sigma^{abc}= \sigma^{[a} \tilde{\sigma}^{b}\sigma^{c]}$, 
$\tilde{\sigma}^{abc}=   \tilde{\sigma}^{[a}  \sigma^{b} \tilde{\sigma}^{c]}$, 
{\it etc.}. 
Finally, 
\begin{eqnarray}
\label{rhC=hC+hC+}
type~ IIA: \qquad & \hat{C} =\hat{C}_1 \oplus \hat{C}_3  \oplus 
\hat{C}_5 \oplus \hat{C}_7  \oplus \hat{C}_9 \; , & \qquad 
\\ 
\label{rC=C+C+}
type~ IIB: \qquad & C=C_0 \oplus C_2 \oplus C_4  \oplus C_6  \oplus C_8 
\oplus C_{10} 
\;    
& \qquad 
\end{eqnarray}
denote the formal sums  of all type IIA (odd) and all type IIB (even)  
RR superforms 
\begin{eqnarray}
\label{rCIIA}
type~ IIA: \qquad & \hat{C}_{2n+1}
={1 \over (2n+1)!}
d\hat{Z}^{ {\hat{M}}_{2n+1}} \wedge \ldots \wedge
d\hat{Z}^{ {\hat{M}}_1}
\hat{C}^{(2n+1)}_{ {\hat{M}}_1\ldots
 {\hat{M}}_{2n+1}}(\hat{Z})\; ,
\\ \label{rCIIB} 
type~ IIB: \qquad &
C_{2n}
={1 \over 2n!}
dZ^{ {M}_{2n}} \wedge \ldots \wedge
dZ^{ {M}_1}C^{(2n)}_{ {M}_1\ldots  {M}_{2n}}(Z) \; .
\end{eqnarray}

\subsection{Isometries and underlying 
superspace ${\cal M}^{(11|32)}$}

The coordinates associated with the isometry directions 
of type IIA and type IIB superspaces are denoted by  
$\hat{z}$ and $y$, respectively. 
The existence of such isometries provides  the  necessary condition for 
the existence of a T--duality map. 
Then we identify {\sl all} 
the remaining superspace coordinates of {\sl curved}  
type IIA and type IIB superspace, i.e. 
  \begin{eqnarray}
\label{MIIA}
 type~IIA: & \qquad     {\cal M}_{IIA}^{(10|32)} \; : & \quad   
    \hat{Z}^{ {\hat{M}}} = 
 (\tilde{{Z}}{}^{ {\tilde{M}}}, \hat{z})\; ,
\\ \label{MIIB} 
  type~IIB: & \qquad   {\cal M}_{IIB}^{(10|32)} \; :  & \quad   
{Z}^{ {{M}}}= 
 (\tilde{Z}^{\tilde{M}}, y )\; . 
\end{eqnarray}
In other words, we assume that the intersection of curved type IIA and 
type IIB superspaces, ${\cal M}_{IIA}^{(10|32)}$ and 
${\cal M}_{IIB}^{(10|32)}$, defines some $D=9$, $N=2$ superspace 
${\cal M}^{(9|32)}$ 
\begin{eqnarray} 
\label{M9}
& {\cal M}_{IIA}^{(10|32)} \cap  {\cal M}_{IIB}^{(10|32)} = 
 {\cal M}^{(9|32)} \;  \\\   
\label{ZM9} {\cal M}^{(9|32)} \; : 
\tilde{Z}^{ \tilde{M}}
      \equiv \left(
          \tilde{X}^{\tilde{m}}, 
       {\theta}^{\mu}\right) \; , 
\\ 
\nonumber &
 {\tilde{m}}= 0, \ldots , 8 \; , \qquad 
 {{\mu}}= 1, \ldots , 32 \; . \qquad 
\end{eqnarray}
This implies that {\sl we 
consider  T--duality as an operation acting on differential 
forms in superspace rather than on the superspace coordinates}. 
Such a possibility is guarantied by (super)diffeomorphism invariance 
of (superspace super)gravity, {\it i.e.} by 
its  gauge symmetry under arbitrary 
changes of local coordinate system (in superspace)  
\footnote{(Super)diffeomorphism invariance 
 allows one to replace any coordinate transformations by the equivalent  
transformations  of the supergravity (super)fields (see, {\it e.g.},  
\cite{BdAI} and refs. therein).}. However, such 
`picture changing' allows 
to breakthrough the problems which hampered the way to superfield 
T--duality rules (see e.g. \cite{simon2}). 

Moreover, this point of view 
makes transparent that type IIA and type IIB  theories 
with isometries $\partial_{\hat{z}}$ and $\partial_{y}$
can be defined 
on the hypersurfaces $\hat{z}=0$ and $y=0$ of an {\it underlying superspace} 
${\cal M}^{(11|32)}$ with $11$ bosonic and $32$ fermionic coordinates,  
\begin{eqnarray}
\label{M11}
  & {\cal M}^{(11|32)} \; : \quad   
   (\tilde{{Z}}{}^{ {\tilde{M}}}, \; y, \; \hat{z}\;) \; .
\end{eqnarray}

The notion of the 
underlying superspace ${\cal M}^{(11|32)}$ will be useful to obtain 
the superfield T--duality rules. An unholonomic basis of 
${\cal M}^{(11|32)}$ should contain $11$ bosonic superforms 
which could be chosen 
as `mostly IIA', 
\begin{eqnarray}
\label{M11EbA}
(\hat{E}^{\hat{a}}, {E}^{*})\equiv  
(\hat{E}^{\tilde{a}}, \hat{E}^{\#}, E^* ) \; , 
\end{eqnarray}
or as `mostly IIB', 
\begin{eqnarray}
\label{M11EbB}
 (E^a, \hat{E}^{\#})\equiv  (E^{\tilde{a}}, E^{*}, \hat{E}^{\#} ) \; .
\end{eqnarray}
This basis should also contain 32 fermionic supervielbein forms, as the 
underlying superspace has 32 fermionic directions. 
It is convenient to use either IIA forms (\ref{fEIIA}) or IIB forms 
(\ref{fEIIB}).

\subsection{On chirality and fermionic coordinates.}

The possibility of identification of the fermionic coordinates 
of {\sl curved} type IIA and type IIB superspaces, 
$\hat{\theta}^{{\hat{\mu}}}={\theta}^{{{\mu}}}$,  used 
before Eqs. (\ref{MIIA}), (\ref{MIIB}), would not seem surprising if 
one remembers that the fermionic coordinates of a general 
{\sl curved} superspace do not carry any chirality. 
In contradistinction to the case of flat superspace,  
the indices $\hat{{\mu}}$ and ${\mu}$   
are {\sl not} the spinor indices of the Lorentz group;  
$\hat{\theta}^{{\hat{\mu}}}$ and ${\theta}^{{{\mu}}}$ are 
rather transformed by the general superdiffeomorphism symmetry. 
On the other hand, the {\sl chirality}, which is used  to distinguish the 
type IIA and type IIB 
cases,  is defined through the projectors constructed from $D=10$ 
Dirac matrices and, thus, is related to the 
concept of $SO(1,9)$ Lorentz group spinor representation. 
Hence, in curved superspace, {\sl the chirality is a characteristic   
of the fermionic supervielbein 1--forms}, Eqs.   
(\ref{fEIIA}), (\ref{fEIIB}), which do carry $SO(1,9)$ spinor indices. 

In both type IIA and type IIB they  can be considered as 
a pair of fermionic forms carrying Majorana-Weyl $SO(1,9)$ spinor indices
${\a}= 1, \ldots , 16$.
As there is no charge conjugation matrix in the  Majorana--Weyl 
spinor representation of $SO(1,9)$, there is no way to lower or to raise 
the spinor indices. 
Thus the  chirality can be identified with the position  
of the spinor indices of the fermionic supervielbein forms. 
The type IIB theory has both fermionic supervielbeins of 
the same chirality (\ref{fEIIB}) and is chiral, while 
the type IIA theory has fermionic supervielbein forms of opposite chiralities
(\ref{fEIIA}) and is nonchiral. 

However, this does not imply 
different properties of the fermionic coordinates of the curved type IIA and 
type IIB superspaces.  
Only in the flat superspace limits, when one takes the fermionic 
supervielbein to be  derivatives of the fermionic coordinates, 
the chiral structure, together  
with the definite spinor representation of the Lorentz group,   
becomes adjusted to the fermionic coordinates of flat superspace.

\subsection{Isometries and differential forms in superspace}

Arbitrary type IIA (type IIB) super--q--forms  
\begin{eqnarray}\label{IIA-Om}
\hat{\Omega}_q &=& {1\over q!} d\hat{Z}^{M_q} \wedge \ldots \wedge 
d\hat{Z}^{M_1} 
\hat{\Omega}_{M_1\ldots M_q}(\hat{Z}) \; , \quad 
\qquad  \nonumber \\ 
 \Omega_q &=& {1\over q!} dZ^{M_q} \wedge \ldots \wedge dZ^{M_1} 
\Omega_{M_1\ldots M_q}(Z) \; \qquad 
\end{eqnarray}
allow the decomposition 
\begin{eqnarray}\label{IIA-Omd} 
\hat{\Omega}_q = \hat{\Omega}^{(-)}_q + i_{\hat{z}}\hat{\Omega}\wedge d\hat{z} 
\; ,  \qquad 
 {\Omega}_q = {\Omega}^{(-)}_q + i_{y}\hat{\Omega}\wedge dy 
\; , 
\end{eqnarray}
into parts which, respectively, 
contain and do not contain the differentials of the isometry 
coordinate, $d\hat{z}$ or $dy$,  
\begin{eqnarray}\label{IIA-iOmd} 
& i_{\hat{z}}\hat{\Omega}_q := {1\over (q-1)!} d\tilde{Z}^{\tilde{M}_{q-1}} 
\wedge \ldots \wedge 
d\tilde{Z}^{\tilde{M}_1} 
\hat{\Omega}_{\hat{z} \tilde{M}_1\ldots \tilde{M}_{q-1}}(\tilde{Z}) 
\; , 
\\ 
\label{IIA-Omd-}
& \hat{\Omega}^{(-)}_q := {1\over q!} d\tilde{Z}^{\tilde{M}_{q}} 
\wedge \ldots \wedge 
d\tilde{Z}^{\tilde{M}_1} 
\hat{\Omega}_{\tilde{M}_1\ldots \tilde{M}_{q}}(\tilde{Z}) 
\; ;  \\ \nonumber \\  
\label{IIB-iOmd}
& i_{y}\hat{\Omega}:=
{1\over (q-1)!} d\tilde{Z}^{\tilde{M}_{q-1}} 
\wedge \ldots \wedge 
d\tilde{Z}^{\tilde{M}_1} 
\Omega_{y\tilde{M}_1\ldots \tilde{M}_{q-1}}(\tilde{Z}) 
\; , \\  
\label{IIB-Omd-}
& {\Omega}^{(-)}_q:=
{1\over q!} d\tilde{Z}^{\tilde{M}_{q}} 
\wedge \ldots \wedge 
d\tilde{Z}^{\tilde{M}_1} 
\Omega_{\tilde{M}_1\ldots \tilde{M}_{q}}(\tilde{Z}) 
\; .
\end{eqnarray}
The conditions of isometry, that is of 
independence of all superfields on the coordinate $\hat{z}$ resp.  $y$, 
have been already indicated in Eqs. (\ref{IIA-iOmd})--(\ref{IIB-Omd-}).

We find convenient to use the supervielbein forms adapted to the 
isometry, i.e. to assume that they obey the 
superfield generalization  of the Kaluza--Klein ansatz 
\cite{ansatz}.  
For the bosonic supervielbein forms it implies the separation of 
the one tangent (super)space bosonic direction, denoted by 
$\hat{a}=\#$ for type IIA and $a= *$ for type IIB,  
 \begin{eqnarray}\label{EaIIAd}
& type \; IIA \; : & \hat{E}^{\hat{a}}= (\hat{E}^{\tilde{a}}, \hat{E}^{\#}) 
\; , 
\\ \label{EaIIBd} & type \; IIB \; : & 
E^a = ({E}^{\tilde{a}}, {E}^{*}) \; , 
\\ \nonumber && \tilde{a}=0,1, \ldots , 8 \; ,
\end{eqnarray}
and the assumption that the subsets of nine bosonic one-forms 
$\hat{E}^{\tilde{a}}$ and ${E}^{\tilde{a}}$ are defined on the 
nine--dimensional superspace ${\cal M}^{(9|32)}$ (\ref{M9}) 
\begin{eqnarray}\label{KKEaA}
& type \; IIA \; : & 
\hat{E}^{ {\tilde{a}}}= \hat{E}^{ {\tilde{a}}(-)}=
d\tilde{Z}^{\tilde{M}} \hat{E}^{ {\tilde{a}}}_{\tilde{M}} (\tilde{Z}) \qquad 
\Rightarrow \qquad \; 
i_{\hat{z}}\hat{E}^{ {\tilde{a}}}\equiv\hat{E}_{\hat{z}}^{ {\tilde{a}}} =0 
\; , \qquad 
\\ \label{KKEaB}
& type \; IIB  \; : & 
{E}^{ {\tilde{a}}}= {E}^{ {\tilde{a}}(-)} = 
d\tilde{Z}^M  {E}^{ {\tilde{a}}}_M(\tilde{Z}) \;  \qquad 
\Rightarrow  \qquad  \; 
i_{y}{E}^{ {\tilde{a}}}\equiv {E}_{y}^{ {\tilde{a}}} =0  \; . \qquad 
\end{eqnarray}
We find convenient to use the notation  
$i_{\hat{z}}\hat{E}^{{\tilde{a}}}$ and $i_{y}{E}^{ {\tilde{a}}}$ instead of 
$\hat{E}_{\hat{z}}^{{\tilde{a}}}$ and ${E}_{y}^{ {\tilde{a}}}$. 

Thus $d\hat{z}$ and $dy$ differentials appear only in the bosonic superforms 
$\hat{E}^{\#}$ and $E^*$ respectively, 
\begin{eqnarray}\label{KKE9A}
& type \; IIA  \; : & 
\hat{E}^{\#}= 
\hat{E}^{\# (-)}+ i_{\hat{z}}\hat{E}^{\#} d\hat{z} \; , 
\quad 
\nonumber \\ && 
\hat{E}^{\# (-)}=d\tilde{Z}^{\tilde{M}} \hat{E}_{\tilde{M}}^{\#}
(\tilde{Z})\; , \quad  
 i_{\hat{z}}\hat{E}^{\#} = 
\hat{E}_{\hat{z}}^{\#}(\tilde{Z})\; , \qquad 
\\ \label{KKE9B}
& type \; IIB  \; : & 
{E}^{*}= {E}^{*(-)} + i_yE^* dy \; , \quad 
 \nonumber \\ && 
{E}^{*(-)}= 
d\tilde{Z}^{\tilde{M}} {E}_{\tilde{M}}^{*}(\tilde{Z}) \; , \quad 
i_{y}{E}^{*}=E_y^*(\tilde{Z})  \; , \qquad 
\end{eqnarray}
but all the superfields in the decompositions (\ref{KKE9A}), (\ref{KKE9B}) 
depend only on the coordinates $\tilde{Z}^{\tilde{M}}$  of the 
nine--dimensional superspace (\ref{M9}).

The superfield generalization of the Kaluza--Klein ansatz 
\cite{ansatz} allows the appearance of $d\hat{z}$ ($dy$) terms in the 
fermionic supervielbein forms (\ref{fEIIA}), (\ref{fEIIB}) as well.  
For superspace calculations it is convenient to redefine  the 
decomposition (\ref{IIA-Omd}) of the fermionic forms  by using 
the bosonic supervielbein forms $\hat{E}^{ {\#}}$ and ${E}^{ {*}}$ 
instead of $d\hat{z}$ ($dy$),  
 \begin{eqnarray}\label{EfIIA}
& type\;  IIA : &      \hat{E}^{{\a}1}=
      \hat{E}^{{\a}1(-)} + 
i_{\hat{z}}\hat{E}^{{\a}1} d\hat{z}  = 
\hat{E}^{{\a}1[-]} + \hat{E}^{\#} {i_{\hat{z}}\hat{E}^{{\a}1}\over 
i_{\hat{z}}\hat{E}^{\#}} \; , 
\qquad   \nonumber \\ && 
\hat{E}_{\a}^2=
      \hat{E}_{\a}^{2(-)} + i_{\hat{z}}\hat{E}_{\a}^2 d\hat{z} = 
\hat{E}_{\a}^{2[-]} + \hat{E}^{\#} {i_{\hat{z}}\hat{E}_{\a}^2\over 
i_{\hat{z}}\hat{E}^{\#}} \; ,
\\ \label{EfIIB}
& type \; IIB : &     
{E}^{ {{\a}}1}= {E}^{ {{\a}}1(-)} + i_y {E}^{ {{\a}}1}dy=
 {E}^{ {{\a}}1[-]} + E^* {i_y {E}^{ {{\a}}1}\over i_y E^*} 
      \; , \qquad   \nonumber \\ &&
     {E}^{ {{\a}}2}= {E}^{ {{\a}}2(-)} + i_y {E}^{ {{\a}}2} dy=
 {E}^{ {{\a}}2[-]} + E^* {i_y {E}^{ {{\a}}2}\over i_y E^*} \; .        
\end{eqnarray}
Clearly, the relations between the forms 
$\hat{E}^{{\a}1[-]}$, $\hat{E}_{\a}^{2[-]}$, 
${E}^{ {{\a}}1,2[-]}$ and the forms 
$\hat{E}^{{\a}1(-)}$, $\hat{E}_{\a}^{2(-)}$, 
${E}^{ {{\a}}1,2(-)}$ of the standard decomposition 
(\ref{IIA-iOmd})--(\ref{IIB-Omd-}) read 
 \begin{eqnarray}\label{EfIIA-}
type\;  IIA : &      
\hat{E}^{ {{\a}1}(-)}= 
\hat{E}^{{\a}1[-]} +
     \hat{E}^{\#(-)} {i_{\hat{z}}\hat{E}^{{\a}1}\over 
i_{\hat{z}}\hat{E}^{\#}} \; , 
\qquad 
\hat{E}^{2(-)}_{ {{\a}}} = \hat{E}^{2[-]}_{ {{\a}}} +             
     \hat{E}^{ {\#}(-)} {i_{\hat{z}}\hat{E}_{\a}^2\over 
i_{\hat{z}}\hat{E}^{\#}}\; , \qquad 
\\ \label{EfIIB-}
 type \; IIB : &  
{E}^{{\a}1(-)}=
                  E^{ {{\a}}1[-]} +
     {E}^{ {*} (-)} {i_y {E}^{ {{\a}}1}\over i_y E^*}  \; , 
\qquad 
{E}^{{\a}2(-)}=
                  E^{ {{\a}}2[-]} +
     {E}^{ {*} (-)} {i_y {E}^{ {{\a}}2}\over i_y E^*}  \; . 
\end{eqnarray}
Such a decomposition is useful as well for the spin connections, 
\begin{eqnarray}\label{wabIIA-}
& \hat{w}^{\hat{a}\hat{b}}=  \hat{w}^{\hat{a}\hat{b}(-)}
+ d\hat{z} \hat{w}_{\hat{z}}^{\hat{a}\hat{b}}(\tilde{Z})
= \hat{w}^{\hat{a}\hat{b}[-]} + 
{\hat{E}^{\#}\over i_{\hat{z}} \hat{E}^{\#}} 
 \hat{w}_{\hat{z}}^{\hat{a}\hat{b}}(\tilde{Z})\; , \qquad 
 \nonumber \\ 
& \hat{w}^{\hat{a}\hat{b}[-]}= 
\hat{w}^{\hat{a}\hat{b}(-)}- {\hat{E}^{\# (-)}\over i_{\hat{z}} \hat{E}^{\#}} 
 \hat{w}_{\hat{z}}^{\hat{a}\hat{b}}(\tilde{Z})\equiv 
\hat{E}^{\tilde{a}(-)} 
\hat{w}_{\tilde{a}}^{\hat{a}\hat{b}}(\tilde{Z})
+ \hat{E}^{\alpha 1[-]} 
\hat{w}_{\alpha 1}^{\hat{a}\hat{b}}(\tilde{Z})
+ \hat{E}_{\alpha}^{2[-]} 
\hat{w}^{\alpha}_{2}{}^{\hat{a}\hat{b}}(\tilde{Z}) 
\; , \qquad 
\\ \label{wabIIB-}
& {w}^{{a}{b}} =  
{w}^{{a}{b}(-)} + dy
 {w}_{y}^{{a}{b}}(\tilde{Z})
= {w}^{{a}{b}[-]} + {{E}^{*}\over i_y{E}^{*}} 
 {w}_{y}^{{a}{b}}(\tilde{Z})
\; , \qquad 
\nonumber \\ 
& {w}^{{a}{b}[-]}= {w}^{{a}{b}(-)} - {{E}^{*}\over i_y{E}^{*}} 
 {w}_{y}^{{a}{b}}(\tilde{Z})= 
{E}^{\tilde{a}(-)} 
{w}_{\tilde{a}}^{{a}{b}}(\tilde{Z})
+  
{E}^{\alpha 1[-]} 
{w}_{\alpha 1}^{{a}{b}}(\tilde{Z})
+ {E}^{\alpha 2[-]} 
{w}_{\alpha 2}^{{a}{b}}(\tilde{Z})\; . \qquad 
\end{eqnarray}

The use of supervielbein forms adapted to the isometry allows us to 
write the superfield T--duality rules in a compact form.

\setcounter{equation}0
\section{R\'esum\'e of bosonic T-duality rules}

Let us begin by a brief r\'esum\'e of the well--known bosonic results. The 
T--duality rules for NS--NS fields (Buscher rules) \cite{buscher} 
have the simplest form in the string frame
\begin{eqnarray}\label{TrAS}
String\; frame \; :  
& g^{(s)}_{yy}={1\over \hat{g}^{(s)}_{\hat{z}\hat{z}}}\; ,
\qquad
{g}^{(s)}_{\tilde{ {m}}y}= 
{1\over \hat{g}_{\hat{z}\hat{z}}}
\hat{B}_{\hat{z}\tilde{ {m}}}\; ,
\qquad \\ \label{TrulesA-S}
& g^{(s)}_{\tilde{ {m}}\tilde{ {n}}} =
\hat{g}^{(s)}_{\tilde{ {m}}\tilde{ {n}}}
+ {1\over \hat{g}^{(s)}_{\hat{z}\hat{z}}}
\left ( {\hat{B}}_{\tilde{ {m}} \hat{z}}
{\hat{B}}_{\tilde{ {n}} \hat{z}}
- \hat{g}^{(s)}_{\tilde{ {m}} \hat{z}}
\hat{g}^{(s)}_{\tilde{ {n}} \hat{z}}\right)\; . \qquad
\\ 
\label{TdilatBS}
& e^{2\Phi} =- {e^{2\hat{\Phi}}\over \hat{g}^{(s)}_{\hat{z}\hat{z}}}\; , \qquad
\\ 
\label{T-BbS}
& B_{\tilde{{ {m}}}\tilde{{ {n}}}} =
\hat{B}_{\tilde{{ {m}}}\tilde{{ {n}}}}
+ {1\over \hat{g}^{(s)}_{\hat{z}\hat{z}}}
\left ( \hat{g}^{(s)}_{\tilde{{ {m}}}\hat{z}}
\hat{B}_{\tilde{ {n}} \hat{z}} -
\hat{g}^{(s)}_{\tilde{ {n}} \hat{z}}
\hat{B}_{\tilde{ {m}} \hat{z}}
\right), \nonumber \\
& B_{y\tilde{ {m}}} =
 {1\over \hat{g}^{(s)}_{\hat{z}\hat{z}}} 
\hat{g}^{(s)}_{\tilde{ {m}} \hat{z}}\; .
\end{eqnarray}
In \cite{simon} they were rederived from the 
relation between 
type IIA and type IIB D--brane actions in purely bosonic supergravity 
background (i.e. in the background of the  bosonic fields 
of the supergravity multiplets). 
Moreover, in \cite{simon} the rules for the RR gauge fields 
were obtained as well. They are 
\begin{eqnarray}\label{0C0=}
C^{(0)}&=& \hat{C}^{(1)}_{\hat{z}} \; ,
\\ \label{0C2ny=}
C^{(2n)}_{y\tilde{m}_1\ldots \tilde{m}_{2n-1}} &=&
\hat{C}^{(2n-1)}_{\tilde{m}_1\ldots \tilde{m}_{2n-1}} +
(2n-1)
\hat{C}^{(2n-1)}_{\hat{z}[\tilde{m}_1\ldots \tilde{m}_{2n-2}}
\hat{g}_{\tilde{m}_{2n-1}]\hat{z}}/ {\hat{g}_{\hat{z}\hat{z}}} \; ,
\\ \label{0C2n-=}
C^{(2n)}_{\tilde{m}_1\ldots \tilde{m}_{2n}}&=&
\hat{C}^{(2n+1)}_{\hat{z}\tilde{m}_1\ldots \tilde{m}_{2n}} +
2n \hat{C}^{(2n-1)}_{[\tilde{m}_1\ldots \tilde{m}_{2n-1}}
\hat{B}_{\tilde{m}_{2n}]\hat{z}} + \nonumber \\
 &+& 2n(2n-1) \hat{C}^{(2n-1)}_{\hat{z}[\tilde{m}_1\ldots
\tilde{m}_{2n-2}}
\hat{B}_{\hat{z}\tilde{m}_{2n-1}}\hat{g}_{\tilde{m}_{2n}]\hat{z}}
/{\hat{g}_{\hat{z}\hat{z}}} \; , \\ 
\nonumber for \; n=1,2,3,4 \; , 
\\ \label{0C10=} 
C^{(10)}_{y\tilde{m}_1\ldots \tilde{m}_{9}} &=&
\hat{C}^{(9)}_{\tilde{m}_1\ldots \tilde{m}_{9}} +
9 
\hat{C}^{(9)}_{\hat{z}[\tilde{m}_1\ldots \tilde{m}_{8}}
\hat{g}_{\tilde{m}_{9}]\hat{z}}/ {\hat{g}_{\hat{z}\hat{z}}} \; .
\end{eqnarray}

In the Einstein frame, where the metric is redefined as  
\begin{eqnarray}\label{Eig=g}
& \hat{g}^{(s)}_{ {\hat{m}} {\hat{n}}}= 
e^{{\hat{\Phi}\over 2}}
\hat{g}_{ {\hat{m}} {\hat{n}}}\; , \qquad 
{g}^{(s)}_{ {{m}} {{n}}}= 
e^{{\Phi\over 2}}
{g}_{ {{m}} {{n}}}\; , 
\end{eqnarray}
the NS--NS T--duality rules read 
\begin{eqnarray}\label{TrA}
& e^{{\Phi\over 2}}g_{yy}={1\over 
e^{{\hat{\Phi}\over 2}}\hat{g}_{\hat{z}\hat{z}}}\; ,
\qquad
e^{{\Phi\over 2}} {g}_{\tilde{ {m}}y}= 
{1\over e^{{\hat{\Phi}\over 2}}
\hat{g}_{\hat{z}\hat{z}}}
\hat{B}_{\hat{z}\tilde{ {m}}}\; , & 
\qquad \\ \label{TrulesA-}
& e^{{\Phi\over 2}} 
g_{\tilde{ {m}}\tilde{ {n}}} = e^{{\hat{\Phi}\over 2}}
\hat{g}_{\tilde{ {m}}\tilde{ {n}}}
+ {1\over e^{{\hat{\Phi}\over 2}} \hat{g}_{\hat{z}\hat{z}}}
\left ( {\hat{B}}_{\tilde{ {m}} \hat{z}}
{\hat{B}}_{\tilde{ {n}} \hat{z}}
- e^{\hat{\Phi}}\hat{g}_{\tilde{ {m}} \hat{z}}
\hat{g}_{\tilde{ {n}} \hat{z}}\right)\; , & \qquad
\\ \label{TdilatB}
& e^{2\Phi} =- {e^{2\hat{\Phi}}\over e^{{\hat{\Phi}\over 2}}
\hat{g}_{\hat{z}\hat{z}}}\; , & \qquad
\\ 
\label{T-Bb}
& B_{\tilde{{ {m}}}\tilde{{ {n}}}} =
\hat{B}_{\tilde{{ {m}}}\tilde{{ {n}}}}
+ {1\over \hat{g}_{\hat{z}\hat{z}}}
\left ( \hat{g}_{\tilde{{ {m}}}\hat{z}}
\hat{B}_{\tilde{ {n}} \hat{z}} -
\hat{g}_{\tilde{ {n}} \hat{z}}
\hat{B}_{\tilde{ {m}} \hat{z}}
\right), & \quad \nonumber \\ 
& B_{y\tilde{ {m}}} =
 {1\over \hat{g}_{\hat{z}\hat{z}}} 
\hat{g}_{\tilde{ {m}} \hat{z}}\; . & \qquad 
\end{eqnarray}
Clearly, the rules for NS--NS two--forms, Eqs. (\ref{T-Bb}), as well as 
the RR T--duality rules (\ref{0C0=}), (\ref{0C2ny=}), (\ref{0C2n-=}), 
(\ref{0C10=})  keep the same form in the Einstein frame. 

\section{T--duality rules for bosonic  superforms} 
\setcounter{equation}0

The T--duality rules for the 
bosonic superforms of type IIA and type IIB supergravity, 
(\ref{bEIIA}), 
(\ref{0B2IIA}), (\ref{rCIIA})
and 
(\ref{bEIIB}),  (\ref{0B2IIB}), (\ref{rCIIB}) 
can be derived from the study of the relations between the  
complete $\kappa$--symmetric  super--Dp--brane actions, i.e.  
by the superfield generalization of the method proposed in \cite{simon}.
We will describe these calculations in a longer paper. 
Here, instead, we will perform a straightforward superfield 
(more precisely, superform) generalization of the pure bosonic 
rules from \cite{simon}, which reproduces exactly the same results. 
To this end
\begin{itemize}
\item one rewrites (\ref{TrA})--(\ref{T-Bb}),   
(\ref{0C0=})--(\ref{0C10=})
in differential form notations, using, in particular, the bosonic 
vielbein $e^{ {a}}= dx^{ {m}}
e^{ {a}}_{ {m}}(x)$ instead of the metric 
$g_{ {m} {n}}(x)= e^{ {a}}_{ {m}}e_{ {a} {n}}$. Note that, e.g.,  
\begin{eqnarray}\label{g/gzz}
{\hat{g}_{\tilde{m}\hat{z}}\over \hat{g}_{\hat{z}\hat{z}}} \equiv  
 {\hat{e}_{\hat{a}\tilde{m}}\hat{e}^{\hat{a}}_{\hat{z}} 
\over \hat{e}_{\hat{b}\hat{z}}\hat{e}^{\hat{b}}_{\hat{z}}} = 
 {\hat{e}^{\#}_{\tilde{m}} 
\over \hat{e}^{\#}_{\hat{z}}} \; , 
\end{eqnarray}
where the second equality is valid for the frame adapted to the 
isometry, i.e. follows from the usual  
Kaluza--Klein ansatz: 
$\hat{e}^{\tilde{a}}= d\tilde{x}^{\tilde{m}} \hat{e}_{\tilde{m}}^{\tilde{a}}
(\tilde{x})$, ($\tilde{m}=0, \ldots , 8$, $\tilde{a}=0, \ldots , 8$), 
which implies $\hat{e}^{\tilde{a}}_{\hat{z}}=0$ ({\it cf.} (\ref{KKEaA})). 

\item One replaces all  fields by superfields, but assumes independence  
on one superspace bosonic coordinate, 
$\hat{z}=\hat{X}^{9}$ for type IIA  and $y=X^9$ for type IIB superfields, 
which describe  the bosonic isometry directions.   
\end{itemize} 

In such a way, starting from Eqs. (\ref{TrA})--(\ref{T-Bb}), 
one reproduces 
the following {\sl superform generalization of the 
NS--NS T--duality rules}  (\ref{TrA})--(\ref{T-Bb})
\begin{eqnarray}\label{T-EIIII}
e^{{\Phi(\tilde{Z})\over 4}} E^{ {\tilde{a}}(-)} &=& 
e^{{\hat{\Phi}(\tilde{Z})\over 4}}\hat{E}^{ {\tilde{a}}(-)} \; , 
\\ \label{T-Eyz}
e^{{\Phi\over 4}} i_yE^{ {*}}&=& {1\over 
e^{{\hat{\Phi}\over 4}} i_{\hat{z}}\hat{E}^{ {\#}}}\; , 
\\  \label{T-Ey-}
e^{{\Phi\over 4}} E^{ {*}(-)}&=& {i_{\hat{z}} 
\hat{B}_2 \over e^{{\hat{\Phi}\over 4}} i_{\hat{z}}\hat{E}^{ {\#}}} \; , 
\\ \label{T-Phi}
e^{{\Phi}(\tilde{Z})} &=& {e^{\hat{\Phi}(\tilde{Z})}\over 
e^{{\hat{\Phi}\over 4}} i_{\hat{z}}\hat{E}^{ {\#}}} \; , 
\\  \label{T-By}
i_yB_2 &=& {\hat{E}^{ {\#}(-)}\over 
i_{\hat{z}}\hat{E}^{ {\#}}} \; , 
\\  \label{T-B-}
B_2^{(-)}&=& \hat{B}^{(-)}_2 - i_{\hat{z}}\hat{B}_2\wedge 
 {\hat{E}^{ {\#}(-)}\over i_{\hat{z}}\hat{E}^{ {\#}}} \; . 
\end{eqnarray}
Here we have used the decomposition (\ref{IIA-Omd}) 
for type IIA and type IIB NS--NS superforms as well as 
the supervielbein forms adapted to the isometry, Eqs. 
(\ref{EaIIAd})--(\ref{KKE9B}) ({\it i.e.}, 
obeying the superfield Kaluza--Klein ansatz \cite{ansatz}).

In the same manner, writing Eqs. 
(\ref{0C0=}), (\ref{0C2ny=}), (\ref{0C2n-=}), (\ref{0C10=})  in the 
differential form notation and replacing all the forms by superforms 
subject to the  superspace isometry conditions, 
one can arrive at the T--duality rules for 
the Ramond--Ramond superfield potentials, 
\begin{eqnarray}
\label{r0C=}
C^{(0)} &=&  i_{\hat{z}}\hat{C}_1 \; ,
\\ \label{r2nyC=}
i_{y}C_{2n}  &=& - \hat{C}^{(-)}_{2n-1} +
{\hat{E}^{ {\#}(-)}\over i_{\hat{z}}\hat{E}^{ {\#}}}\,\wedge
i_{\hat{z}} \hat{C}_{2n-1}\; , 
\\ \label{r2n-C=}
C^{(-)}_{2n}
&=& i_{\hat{z}}\hat{C}_{2n+1} +
i_{\hat{z}} \hat{B}_2 \wedge
\left(\hat{C}^{(-)}_{2n-1} -  \,
{\hat{E}^{ {\#}(-)}\over i_{\hat{z}}\hat{E}^{ {\#}}}\,
\wedge
i_{\hat{z}} \hat{C}_{2n-1} \right)\, , 
\\ 
\nonumber for \; n=1,2,3,4 \; , \\ 
\label{r10C=}
i_{y}C_{10}  &=& - \hat{C}^{(-)}_{9} +
{\hat{E}^{ {\#}(-)}\over i_{\hat{z}}\hat{E}^{ {\#}}}\,\wedge
i_{\hat{z}} \hat{C}_{9}\; .
\end{eqnarray}

Note that, if we had made use of  supervielbein forms which were not adapted 
to the isometries as in (\ref{EaIIAd})--(\ref{KKE9B}), 
we would have arrived at more complicated expressions for the 
T--duality rules,
\begin{eqnarray}\label{TE-nad}
& e^{{\Phi(\tilde{Z})\over 4}} E^{a(-)}= 
e^{{\hat{\Phi}(\tilde{Z})\over 4}}\hat{E}^{a(-)} + 
\left(e^{{\hat{\Phi}(\tilde{Z})\over 4}} 
{\hat{E}^{a(-)}i_{\hat{z}}\hat{E}_{a}\over \hat{G}_{\hat{z}\hat{z}}}
- e^{-{\hat{\Phi}(\tilde{Z})\over 4}} {i_{\hat{z}} \hat{B}_2 \over 
G_{\hat{z}\hat{z}}}\right)  i_{\hat{z}}E^a  \; , 
\\ \label{TEynad}
& e^{{\Phi\over 4}} i_yE^{a}= - {i_{\hat{z}}\hat{E}^{a}
\over \hat{G}_{\hat{z}\hat{z}}}\; , \quad 
e^{2{\Phi}} = - {e^{2\hat{\Phi}(\tilde{Z})}\over 
e^{{\hat{\Phi}\over 2}}\hat{G}_{\hat{z}\hat{z}}}\; , \quad 
i_yB_2 =  {\hat{E}^{a(-)} i_{\hat{z}}\hat{E}_{a}
\over 
\hat{G}_{\hat{z}\hat{z}}}\; , \quad 
\\ \label{TBnad}
& B_2^{(-)} = \hat{B}^{(-)}_2 - i_{\hat{z}}\hat{B}_2\wedge 
 {\hat{E}^{a(-)} i_{\hat{z}}\hat{E}_{a}
\over \hat{G}_{\hat{z}\hat{z}}} \; ,  
\end{eqnarray}
where 
\begin{eqnarray}\label{Gzz}
\hat{G}_{\hat{z}\hat{z}}\equiv i_{\hat{z}}\hat{E}^{a}i_{\hat{z}}\hat{E}_{a} 
\; .
\end{eqnarray}
These coincide with the rules from Ref. \cite{kulik} up to the dilaton 
factor, the discrepancy comes from the fact that Ref. \cite{kulik} deals 
with supervielbeins {\sl in the string frame}  
\begin{eqnarray}\label{Estr}
 & type \; IIA \; :  \quad  
({\hat{\cal E}}^{\hat{a}}, {\hat{\cal E}}^{\alpha 1}, 
{\hat{\cal E}}_{\alpha}^{2}) = & 
(e^{{\hat{\Phi}\over 4}} {\hat{E}}^{\hat{a}}, e^{{\hat{\Phi}\over 8}}
{\hat{E}}^{\alpha 1}, e^{{\hat{\Phi}\over 8}} {\hat{E}}_{\alpha}^{2})\; , 
 \qquad  \nonumber \\ & 
type \; IIB \; : \quad  
({{\cal E}}^{\hat{a}}, {{\cal E}}^{\alpha 1}, 
{{\cal E}}^{\alpha 2}) = & 
(e^{{{\Phi}\over 4}} {{E}}^{\hat{a}}, e^{{{\Phi}\over 8}}
{{E}}^{\alpha 1}, e^{{{\Phi}\over 8}} {{E}}^{\alpha 2})\; . \qquad 
\end{eqnarray}

Below we show that the T-duality transformation rules for fermionic 
supervielbein forms can  be derived by using the 
rules for the bosonic forms together with the supergravity 
constraints. These fermionic T--duality rules also coincide with the 
ones from \cite{kulik} after transformation to the string frame. 

However, our approach also allows one to derive the  T--duality 
transformation rules for the 
RR superform potentials, Eqs. (\ref{r0C=})--(\ref{r10C=}).   
For general supervielbeins which are not adapted to the isometry they 
read
\begin{eqnarray}
\label{r0C=nad}
C^{(0)} &=&  i_{\hat{z}}\hat{C}_1 \; ,
\\ \label{r2nyC=nad}
i_{y}C_{2n}  &=& - \hat{C}^{(-)}_{2n-1} +
{\hat{E}^{a(-)}i_{\hat{z}}\hat{E}_a
\over \hat{G}_{\hat{z}\hat{z}}}
\, \wedge
i_{\hat{z}} \hat{C}_{2n-1}  \; ,
\\ \label{r2n-C=nad}
C^{(-)}_{2n}
&=& i_{\hat{z}}\hat{C}_{2n+1} +
i_{\hat{z}} \hat{B}_2 \wedge
\left(\hat{C}^{(-)}_{2n-1} - \, {\hat{E}^{a(-)}i_{\hat{z}}\hat{E}_a
\over \hat{G}_{\hat{z}\hat{z}}} \,
\wedge
i_{\hat{z}} \hat{C}_{2n-1} \right)\, , 
\\ 
\nonumber for \; n=1,2,3,4 \; , 
\\ \label{r10C=nad}
i_{y}C_{10}  &=& - \hat{C}^{(-)}_{9} +
{\hat{E}^{a(-)}i_{\hat{z}}\hat{E}_a
\over \hat{G}_{\hat{z}\hat{z}}}
\, \wedge
i_{\hat{z}} \hat{C}_{9}  \; .
\end{eqnarray} 

Eqs. (\ref{TE-nad})--(\ref{Gzz}), (\ref{r0C=nad})--(\ref{r2n-C=nad})  
possess manifest  Lorentz ($SO(1,9)$) invariance. However, 
one shall keep in mind that they hold for superspaces 
with bosonic isometries. So we prefer to take  
advantage of  supervielbein forms 
adapted to the isometry, Eqs. (\ref{EaIIAd})--(\ref{KKE9B}), 
and to use a simpler and more geometrical form of the T--duality rules, 
Eqs. (\ref{T-EIIII})--(\ref{T-B-}),  (\ref{r0C=})--(\ref{r2n-C=})
\footnote{
If necessary, the generalization to supervielbein forms which are not 
adapted to the isometry can be made quite easily. To this end, 
one should take our equations and replace all the expressions with 
broken ten--dimensional 
Lorentz symmetry by  formally covariant expressions, which are 
equal to the original ones for the adapted supervielbein forms. For instance, 
$$ {\hat{E}^{\# (-)} \over i_{\hat{z}} \hat{E}^{\#}}= 
{\hat{E}^{\# (-)} i_{\hat{z}} \hat{E}^{\#}\over i_{\hat{z}} \hat{E}^{\#}
i_{\hat{z}} \hat{E}^{\#}}= {\hat{E}^{a (-)} i_{\hat{z}} \hat{E}_{a}\over 
\hat{G}_{\hat{z}\hat{z}}}\; , \qquad 
i_{\hat{z}} \hat{E}^{\#} 
\mapsto i_{\hat{z}} \hat{E}^{\#}\delta_{\#}^{\hat{a}}
= i_{\hat{z}} \hat{E}^{\hat{a}}\mapsto  i_{\hat{z}} \hat{E}^{{a}}\; , $$ 
{\it etc.}
Note that in this way one identifies $\hat{a}=a$ to make sense of 
relations like (\ref{TE-nad}).}.

\subsection{Compact form of the T--duality rules for superforms}

For future use it is convenient to 
present the T--duality rules (\ref{T-By}), 
(\ref{T-B-}) as a relation between complete NS--NS superforms 
${{B}}_2$ and   $\hat{{B}}_2$. To this end one uses 
Eqs. (\ref{IIA-Omd}) to rewrite Eq. (\ref{T-B-})  in the following form
\begin{equation}\label{TBsg}
{{B}}_2=  \hat{{B}}_2 -
(dy + i_{\hat{z}} \hat{B}_2) \wedge
\left(d\hat{z}+ {\hat{E}^{ {\#}(-)}
\over i_{\hat{z}}\hat{E}^{ {\#}}}\right) + dy \wedge d\hat{z}\; .   
\end{equation} 
Furthermore, using the T-duality rules (\ref{T-Eyz}), 
(\ref{T-By}) 
(which imply $E^{*(-)}\equiv E^{*}- i_yE^{*} dy
= i_{\hat{z}}\hat{B}_2 i_y E^{*}$, {\it  i.e.}  
$dy+ i_{\hat{z}}\hat{B}_2= E^* /i_y E^{*}$)  
and extracting 
$i_{y}{E}^{ {*}}\, i_{\hat{z}}\hat{E}^{ {\#}}$ in the 
common denominator, one finds 
\begin{equation}
\label{rTdB2} 
B_2 = \hat{B}_2 - \, 
{1\over i_{y}{E}^{ {*}}} \, 
E^{ {*}} \wedge \hat{E}^{ {\# }} \,
{1\over i_{\hat{z}}\hat{E}^{ {\#}}}\, 
+ dy  \wedge d\hat{z} \; .
\end{equation}

In the same manner the T--duality rules for RR superfield potentials, 
Eqs. (\ref{r0C=})--(\ref{r2n-C=}),  can be collected in the following 
compact expression written in terms of the formal sums 
of all type IIB and all type IIA forms, (\ref{rC=C+C+}) and 
(\ref{rhC=hC+hC+}),  
\begin{equation}
\label{rC=} 
C= i_{\hat{z}}\hat{C} + (dy +i_{\hat{z}} \hat{B}_2) \wedge 
\left(\hat{C}^{(-)} -  \,
{\hat{E}^{ {\#}(-)}\over i_{\hat{z}}\hat{E}^{ {\#}}}\, 
\wedge 
i_{\hat{z}} \hat{C} \right)\, .
\end{equation}
Eq. (\ref{rC=}) can be used in its complete form to extract the 
rules for type IIB $2n$ forms $C_{2n}$ up to $C_8$. 
For $C_{10}$ we have only $i_y$ contraction of this equation 
(Eq. (\ref{r10C=nad})). 

The relation inverse to (\ref{rC=}) reads 
\begin{equation}\label{rhC=}
\hat{C} = - i_y C + (d\hat{z}+ i_yB_2)\wedge 
\left(C^{(-)}- i_y C \wedge \,
 {{E}^{ {*}(-)}\over i_{y}{E}^{ {*}}}\, \right)
\, 
\end{equation}
and can be used in its complete form for all type IIA superforms, including 
$\hat{C}_9$.

\section{T--duality rules for fermionic supervielbein forms} 
\setcounter{equation}0

\subsection{Supergravity constraints and spinorial cohomology approach}

In this section we will show that  T--duality rules for the 
fermionic supervielbein forms 
(\ref{EfIIA}), (\ref{EfIIB}) can be derived 
from the rules for the bosonic supervielbein 
(Eqs. (\ref{T-EIIII})--(\ref{T-Phi})) and NS-NS superforms  
(Eqs. (\ref{T-By}), (\ref{T-B-}) or (\ref{rTdB2}))
with the use 
of $D=10$ type IIA and type IIB supergravity constraints \cite{c0}. 

These superspace constraints imply the following expression for the 
bosonic torsion 2--forms 
\begin{eqnarray}
\label{TaIIA=} & IIA \qquad &
\hat{T}^{\hat{a}} 
:= d\hat{E}^{ {\hat{a}}} - 
\hat{E}^{ {\hat{b}}}\wedge 
\hat{w}_{ {\hat{b}}}{}^{ {\hat{a}}}
= - i \hat{E}^{ {\beta}1}\wedge 
\hat{E}^{ {\gamma}1} 
\sigma^{ \hat{a}}_{ {\beta} {\gamma}} - i
\hat{E}_{ {\beta}}^2\, \wedge \, \hat{E}_{ {\gamma}}^2 \,   
\tilde{\sigma}^{\hat{a} {\beta} {\gamma}}
 \; ,
\\ 
\label{TaIIB=} 
& IIB \qquad &
{T}^{ {a}} := d{E}^{ {a}} - 
{E}^{ {b}}\wedge {w}_{ {b}}{}^{ {a}}
= 
- i E^{ {\b}1} \wedge E^{ {\g}1} 
\s^{  {a}}_{ {\b} {\g}} - i 
E^{ {\b}2} \wedge E^{ {\g}2}
\s^{ {a}}_{ {\b} {\g}} 
 \; ,
\end{eqnarray}
and for the NS--NS gauge superfield strength 
\begin{eqnarray}\label{H3IIA}
type \; IIA & \qquad \hat{H}_3:=  d\hat{B}_2   = - i e^{{1\over 2}\hat{\Phi}} 
\hat{E}^{ \hat{a}} 
\wedge (\hat{E}^{ {\beta}1}\wedge \hat{E}^{ {\gamma}1} 
\sigma_{  \hat{a} {\beta} {\gamma}} - 
\hat{E}_{ {\beta}}^2
\wedge \hat{E}_{ {\gamma}}^2  
\tilde{\sigma}_{ \hat{a}}^{ {\beta} {\gamma}}) +    
\\ \nonumber 
& + {e^{{1\over 2}\hat{\Phi}} \over 4} 
\hat{E}^{ \hat{b}} 
\wedge \hat{E}^{ \hat{a}} 
\wedge ( \hat{E}^{ {\beta}1} 
\sigma_{ \hat{a} \hat{b}}
{}_{ {\beta}}{}^{ {\gamma}} 
\hat{\nabla}_{ {\gamma}1}\hat{\Phi} + 
 \hat{E}_{ {\gamma}}^2 
\sigma_{ \hat{a} \hat{b}}
{}_{ {\beta}}{}^{ {\gamma}} 
\hat{\nabla}^{ {\beta}}_2 \hat{\Phi})   +  
{1\over 3!}  \hat{E}^{ \hat{c}} 
\wedge \hat{E}^{ \hat{b}}
\wedge \hat{E}^{ \hat{a}} 
\hat{H}_{\hat{a}\hat{b}\hat{c}}  \; , 
\end{eqnarray}
\begin{eqnarray}\label{H3IIB}
type \; IIB : & {H}_3=  d{B}_2  = 
- i e^{{1 \over 2}\Phi} E^{ {a}} \wedge
(E^{ {\a}1} \wedge E^{ {\b}1} -
E^{ {\a}2} \wedge E^{ {\b}2})
\s_{ {a}~ {\a} {\b}} + 
\\ \nonumber 
& + {1 \over 4} e^{{1 \over 2}\Phi}  
E^{ {b}} \wedge E^{ {a}}  \wedge
   \left( E^{ {\a}1} \nabla_{ {\b}1}\Phi -
   E^{ {\a}2} \nabla_{ {\b}2}\Phi\right)
   (\s_{ {a} {b}})_{ {\a}}^{~ {\b}}
   + {1 \over 3!}
    E^{ {c}}  \wedge
    E^{ {b}}  \wedge
    E^{ {a}} H_{ {a}  {b} {c}}
    \; ,  
\end{eqnarray} 
as well as certain expressions  for the fermionic torsion $2$--forms
($\hat{T}^{ {\beta}1}:= {\cal D} \hat{E}^{ {\beta}1}$, 
 $\hat{T}_{ {\beta}}^2:= {\cal D} \hat{E}_{ {\beta}}^2$ and 
${T}^{ {\beta}1,2}:= {\cal D} {E}^{ {\beta}1,2}$), 
curvatures of the spin connections 
($\hat{R}^{\hat{a}\hat{b}}:= d \hat{w}^{\hat{a}\hat{b}} - 
\hat{w}^{\hat{a}\hat{c}}\wedge \hat{w}_{\hat{c}}{}^{\hat{b}}$ and 
${R}^{{a}{b}}:= d {w}^{{a}{b}} - 
{w}^{{a}{c}}\wedge {w}_{{c}}{}^{{b}}$) and field strengths of the RR 
superforms ($\hat{R}_{2n+2}=d\hat{C}_{2n+1} - \hat{C}_{2n-1} \wedge \hat{H}_3$
and 
$R_{2n+1}:=dC_{2n} - C_{2n-2} \wedge H_3$, see below). For shortness we  
will call {\sl `constraints'}  
all the above mentioned relations (although they include not only  
the proper constraints, see \cite{AlgC,HW84,CGO87}, but also 
their consequences).

The complete set of the superfield T--duality rules should be consistent 
with all the constraints; {\it i.e.} it should map 
the complete set of the type IIA constraints, 
Eqs. (\ref{TaIIA=}), (\ref{H3IIA}), 
{\it etc.}, into  the complete set of 
the 
type IIB constraints,  Eqs. (\ref{TaIIB=}), (\ref{H3IIB}), 
{\it etc.}. 
Clearly, a part of such correspondence should be 
sufficient (and, indeed, is sufficient) to derive  the 
T--duality rules for the fermionic superforms, while the check of the 
correspondence between the remaining constraints promises to be  
the quite  involved  exercise. 

Fortunately, the situation can be simplified drastically by the use of  
theorems about interdependence of the constraints 
(see, {\it e.g.}, \cite{AlgC,HW84,CGO87}). 
The differential form constraints contain a number of spin--tensor and tensor 
relations 
(e.g. $H_{\alpha 1\, \beta 1 \, \gamma 1}=0$, 
$H_{\alpha 1\, \beta 1 \, \gamma 2}=0$, $\ldots$, 
 $H_{\alpha 1\, \beta 1 \, c}= -2i e^{{\Phi\over 2}} \sigma_{c\alpha\beta}$, 
 $\ldots$, $H_{\alpha 1\, b \, c}= 
{1\over 2} e^{{\Phi\over 2}} (\sigma_{bc})_{\alpha}{}^{\beta} 
\nabla_{\beta 1}\Phi$, $\ldots$ in (\ref{H3IIB})) 
which we call {\sl `components'} 
(not to be confused with components of superfields). 
It is convenient to 
classify them by dimension (in energy units and corresponding to the 
lower case indices, e.g. $3/2$, $3/2$, $\ldots$, $2$, $\ldots$, $5/2$, 
$\ldots$ in the above example). Then, 
using the Bianchi identities, $d\hat{H}_3\equiv 0$ and $dH_3\equiv 0$,  
one finds \cite{HW84,CGO87}  that 
all the components of dimension more than $2$ in the constraints 
(\ref{H3IIA}) and (\ref{H3IIB}) can be derived from the lower dimensional 
components of the same equations 
({\it i.e.} coefficients for a basic 3--forms $E^A \wedge E^B\wedge E^C$  
with not more than 
one bosonic supervielbein form $E^a$) 
and the constraints for the 
bosonic torsion two--forms, 
(\ref{TaIIA=}) and (\ref{TaIIB=}), respectively
\footnote{ 
Actually, in such calculations one needs to know as well the constraints for  
the fermionic torsion 2--forms.  
However, they, as well as the expressions for 
the curvatures of spin connections, can be completely restored 
from the constraints for the bosonic torsion two--forms, 
(\ref{TaIIA=}) and (\ref{TaIIB=}), with the use of Bianchi identities  
$${\cal D}\hat{T}^{\hat{a}}= - \hat{E}^{\hat{b}} \wedge \hat{R}_{\hat{b}}{}^
{\hat{a}}\; , \quad  
{\cal D}\hat{T}^{\alpha 1}= - \hat{E}^{\beta 1} \wedge 
\hat{R}^{\hat{b}\hat{a}} {1\over 4} \sigma_{\hat{b}\hat{a}}{}_\beta{}^\alpha
\, , \quad  {\cal D}\hat{T}_{\beta}^2=  \hat{E}_{\beta}^2 \wedge 
\hat{R}^{\hat{b}\hat{a}} {1\over 4} \sigma_{\hat{b}\hat{a}}{}_\alpha{}^\beta 
\, , \quad 
 {\cal D} \hat{R}^{\hat{b}\hat{a}} \equiv 0 \; , $$ and 
$${\cal D}{T}^{{a}}= - {E}^{{b}} \wedge {R}_{{b}}{}^
{{a}}\; , \quad  
{\cal D}{T}^{\alpha 1}= - {E}^{\beta 1} \wedge 
{R}^{{b}{a}} {1\over 4} \sigma_{{b}{a}}{}_\beta{}^\alpha
\, , \quad  
{\cal D}{T}^{\alpha 2}= - {E}^{\beta 2} \wedge 
{R}^{{b}{a}} {1\over 4} \sigma_{{b}{a}}{}_\beta{}^\alpha
\, , \quad 
 {\cal D} {R}^{{b}{a}} \equiv 0 \; . $$ }.

This allows us to search for the T--duality rules for the fermionic 
supervielbein forms by requiring the consistency of the rules for the bosonic 
superforms (Eqs. (\ref{T-EIIII})--(\ref{r10C=})) 
with the 
lower dimensional components of Eqs. (\ref{H3IIA}), (\ref{H3IIB}) and 
the (complete) constraints (\ref{TaIIA=}),  (\ref{TaIIB=}). 
It is convenient to organize this procedure as follows. 
We will study the consistency of the T--duality rules with the complete 
constraints (\ref{TaIIA=})--(\ref{H3IIB}), 
but ignoring in (\ref{H3IIA}), (\ref{H3IIB}) the terms 
${\cal O} (E^a\wedge E^b)$ and ${\cal O} (\hat{E}^{\hat{a}}\wedge 
\hat{E}^{\hat{b}})$, which include more than 
one bosonic supervielbein form. 
Such method is close in spirit to the 
'spinorial cohomology approach' developed recently in \cite{NC} (for 
 a different problem, see also \cite{H03}).

Due to the same reason we can also omit from the consideration all the 
expressions for the fermionic torsions, 
$\hat{T}^{\alpha 1}$, $\hat{T}_{\alpha}^2$ and 
${T}^{\alpha 1,2}$, and for the curvatures,  
$\hat{R}^{\hat{a}\hat{b}}:= d\hat{w}^{\hat{a}\hat{b}}- 
\hat{w}^{\hat{a}\hat{c}}\wedge \hat{w}_{\hat{c}}{}^{\hat{b}}$ 
and  $R^{ab}:= dw^{ab}- w^{ac}\wedge w_c{}^b$.  
Their form can be derived from Eqs. (\ref{TaIIA=})--(\ref{H3IIB}) 
with the use of Bianchi identities (see footnote 5) 
and, hence, their consistency with the 
T--duality rules should be guarantied by the consistency of 
Eqs. (\ref{TaIIA=})--(\ref{H3IIB}). 
The consistency of the T--duality rules with the constraints for 
RR field strength will be discussed in Sec. 5.  

\subsection{Torsion constraints and superfield Kaluza--Klein ansatz} 

First observe that using the superfield Kaluza--Klein ansatz 
for the bosonic supervielbein forms, Eqs. (\ref{KKEaA})--(\ref{KKE9B}), 
in the torsion constraints 
(\ref{TaIIA=}) with $ \hat{a}= \#$ and (\ref{TaIIB=}) with  
$ {a}= *$, one finds (in our notation $\sigma^{\#}=-\sigma_{\#} \equiv 
\sigma^9 \equiv \sigma^*= - \sigma_*$)
\begin{eqnarray} 
\label{chiIIA} & type \; IIA \qquad &
i_{\hat{z}}\hat{E}^{\a 1}= 
{i\over 2} \tilde{\sigma}^{\#  {\a} {\b}}
\hat{\nabla}_{ {\b}1} i_{\hat{z}}\hat{E}^{ {\#}}
\; , \qquad 
i_{\hat{z}}\hat{E}_{\a}^2 
= \, 
 {i\over 2} {\sigma}^{\#}_{ {\a} {\b}} \, 
\hat{\nabla}_2^{ {\b}} i_{\hat{z}}\hat{E}^{ {\#}} \; , 
\qquad 
\\ 
\label{chiIIB} & type \; IIB \qquad &
i_yE^{\a 1}= 
{i\over 2}  \tilde{\sigma}^{\#  {\a} {\b}}
{\nabla}_{ {\b}1}i_{y}{E}^{ {*}}
\; ,  \qquad 
i_y E^{\a 2} = 
{i\over 2}  \tilde{\sigma}^{\#  {\a} {\b}}
{\nabla}_{ {\b}2} i_{y}{E}^{ {*}}\; , \qquad 
\end{eqnarray}
as well as 
\begin{eqnarray} 
\label{wzaz} &
\hat{w}_{\hat{z} \tilde{a}}{}^{\#}= 
- \hat{\nabla}_a i_{\hat{z}}\hat{E}^{ {\#}}\; , 
 \qquad {w}_{y\tilde{a}}{}^{*}= 
- {\nabla}_a i_{y}{E}^{*}\; .
\end{eqnarray} 
Eqs. (\ref{chiIIA}), (\ref{chiIIB}) imply    
that our problem reduces essentially to the search for the 
T--duality rules for $\hat{E}^{\a 1 [-]}$, $\hat{E}_{\a}^{2[-]}$, 
and  $E^{\a 1,2 [-]}$ which, then, will allow to define the rules 
for the fermionic covariant derivative  entering Eqs.  (\ref{chiIIA}), 
(\ref{chiIIB}) (see below). 

Eq. (\ref{TaIIA=}) with $ \hat{a}= \tilde{a}=0,\ldots , 8$ 
and (\ref{TaIIB=}) with  $ {a}= \tilde{a}$ give 
the expressions for type IIA and type IIB representations 
for the bosonic torsion of nine--dimensional superspace (\ref{M9}), 
\begin{eqnarray} 
\label{Ta9IIA} 
& type\;  IIA \; : \quad 
\hat{{\cal D}}^{[-]} \hat{E}^{\tilde{a}(-)} & := 
d\hat{E}^{\tilde{a}(-)} - \hat{E}^{\tilde{b}(-)}\wedge 
\hat{w}^{[-]}{}_{\tilde{b}}{}^{\tilde{a}} = \quad 
\nonumber \\ &&  =
 - i \hat{E}^{{\beta}1[-]}\wedge 
\hat{E}^{ {\gamma}1[-]} 
\sigma^{ \tilde{a}}_{ {\beta} {\gamma}} - i
\hat{E}_{{\beta}}^{2[-]} \wedge \, \hat{E}_{ {\gamma}}^{2[-]} \,   
\tilde{\sigma}^{\tilde{a} {\beta} {\gamma}}
 \; ,  \quad 
\\ 
\label{Ta9IIB} 
& type\;  IIB \; : \quad 
{\cal D}^{[-]} {E}^{\tilde{a}(-)} & := 
d{E}^{\tilde{a}(-)} - {E}^{\tilde{b}(-)}\wedge 
{w}^{[-]}{}_{\tilde{b}}{}^{\tilde{a}} = \quad 
\nonumber \\ &&  =
- i E^{ {\b}1[-]} \wedge E^{ {\g}1[-]} 
\s^{  \tilde{a}}_{ {\b} {\g}} - i 
E^{ {\b}2[-]} \wedge E^{ {\g}2[-]}
\s^{ \tilde{a}}_{ {\b} {\g}} 
 \; , \quad 
\end{eqnarray}
as well as (in the parts proportional to $\hat{E}^{\#}$ and $E^*$) 
specify completely the parts $\hat{w}^{\# \tilde{a}[-]}$ and 
${w}^{* \tilde{a}[-]}$ of the spin connections. Collecting the latter 
result with Eqs. (\ref{wzaz}), we find 
\begin{eqnarray} 
\label{wza} 
& type\;  IIA \; : \quad \hat{w}_{\#}{}^{\tilde{a}} =& 
{1\over i_{\hat{z}}\hat{E}^{\#}} (\hat{E}^{\tilde{b}(-)} 
\hat{w}_{\hat{z}\tilde{b}}{}^{\tilde{a}} 
- 2i \hat{E}^{{\alpha}1[-]}  \sigma^{\tilde{a}}_{ \alpha {\beta}} 
i_{\hat{z}}\hat{E}^{{\beta}1} - \qquad \nonumber \\ && \qquad 
- 2i \hat{E}_{{\alpha}}^{2[-]}
\tilde{\sigma}^{\tilde{a} \alpha {\beta}}
i_{\hat{z}}\hat{E}_{{\beta}}^{2[-]} - 
\hat{E}^{\#}
\hat{\nabla}^{\tilde{a}}i_{\hat{z}}\hat{E}^{\#}\, )\; ,\qquad 
\\ \label{wya} 
& type\;  IIB \; : \quad {w}_{*}{}^{\tilde{a}} =& 
{1\over i_{y}\hat{E}^{*}} ({E}^{\tilde{b}(-)} 
{w}_{y\tilde{b}}{}^{\tilde{a}} 
- 2i {E}^{{\alpha}1[-]}  \sigma^{\tilde{a}}_{ \alpha {\beta}} 
i_{y}{E}^{{\beta}1} - \qquad \nonumber \\ && \qquad 
- 2i {E}^{{\alpha}2[-]}  \sigma^{\tilde{a}}_{ \alpha {\beta}} 
i_{y}{E}^{{\beta}2} - 
{E}^{*}
{\nabla}^{\tilde{a}}i_{y}\hat{E}^{*} \, )\; ,\qquad 
\end{eqnarray}

\subsection{T--duality rules for fermionic forms from 
the supergravity constraints I. General structure from NS-NS constraints }

First, 
let us observe for  future use that Eq. (\ref{TBsg}) allows to derive the 
T--duality rule for the NS--NS gauge superfield strength (\ref{H3IIA}) and 
(\ref{H3IIB}). It is convenient to present them in the form  
\begin{equation}
\label{H3=hH3} 
H_3 = \hat{H}_3 
-  \left(d\hat{z} + 
{\hat{E}^{ {\#}(-)}\over i_{\hat{z}}\hat{E}^{ {\#}}} 
\right)\wedge i_{\hat{z}} \hat{H}_3 \; + 
 \left(dy +  {{E}^{ {*}(-)}\over i_{y}\hat{E}^{ {*}}} \,
\right)\wedge i_y H_3  \; , 
\end{equation}
using the identities $d i_{\hat{z}} \hat{B}_2= - 
i_{\hat{z}} \hat{H}_3\;$,  $di_yB_2= - i_yH_3$ 
implied by the isometry conditions.

On the other hand, taking the exterior derivative of Eq. (\ref{rTdB2}) 
and using the definition of the superspace 
torsion, Eqs.  (\ref{TaIIA=}) and (\ref{TaIIB=}), 
one obtains another (equivalent, but more convenient) 
form  of the T--duality rules (\ref{H3=hH3}), 
\begin{eqnarray}
\label{TdH3} &
H_3 = & \hat{H}_3 
- {1\over i_{y}{E}^{ {*}}\, i_{\hat{z}}\hat{E}^{ {\#}}}
\, E^{ {*}} \wedge \hat{T}^{ {\# }}  
+ {1\over i_{y}{E}^{ {*}}\, i_{\hat{z}}\hat{E}^{ {\#}}}
\, \hat{E}^{ {\#}} \wedge {T}^{ {*}} + \\ 
\nonumber 
 & & + 
{1\over i_{y}{E}^{ {*}}\, i_{\hat{z}}\hat{E}^{ {\#}}}
\left( \hat{E}^{ {\#}} \wedge 
{\cal E}^{ {\tilde{b}}}\wedge 
{w}_{ {\tilde{b}}}{}^{ {*}}
-  {E}^{ {*}} \wedge 
{E}^{ {\tilde{b}}}\wedge \hat{w}_{ {\tilde{b}}}
{}^{ {\#}} +  {E}^{ {*}} \wedge 
\hat{E}^{ {\#}}\wedge d\, 
log|i_{y}{E}^{ {*}}\, i_{\hat{z}}\hat{E}^{ {\#}}|\right) 
\; .
\end{eqnarray}
Substituting 
(\ref{TaIIA=})--(\ref{H3IIB}) 
into (\ref{TdH3}) and taking into account 
Eqs. (\ref{T-EIIII})--(\ref{T-Phi})  
one finds after straightforward algebraic manipulations  
\begin{eqnarray}\label{hH3II2}
 & {}  
\qquad {} 
\qquad  - i e^{{1\over 2}\hat{\Phi}} 
{E}^{ {\tilde{a}}(-)}
& \wedge [ 
({E}^{ {\beta}1}\wedge {E}^{ {\gamma}1} 
\sigma_{  {\tilde{a}} {\beta} {\gamma}} - 
 e^{{1\over 4}(\hat{\Phi}- {\Phi})}\; 
\hat{E}^{ {\beta}1}\wedge \hat{E}^{ {\gamma}1} 
\sigma_{  {\tilde{a}} {\beta} {\gamma}}) \; 
- \\ & & \nonumber 
- \; ({E}^{ {\beta}2}\wedge {E}^{ {\gamma}2} 
\sigma_{  {\tilde{a}} {\beta} {\gamma}} 
+  e^{{1\over 4}(\hat{\Phi}- {\Phi})}\;
\hat{E}_{ {\beta}}^2\wedge \hat{E}_{ {\gamma}}^2  
\tilde{\sigma}_{ {\tilde{a}}}
^{ {\beta} {\gamma}})] + 
\\ & \nonumber 
+ i e^{{1\over 2}{\Phi}}  
({E}^{ {*}}  + e^{{1\over 4}(\hat{\Phi}- {\Phi})}\; 
\hat{E}^{ {\#}}) 
& \wedge [{E}^{ {\beta}1}\wedge {E}^{ {\gamma}1} 
\sigma^{  {\#}}_{ {\beta} {\gamma}} 
- e^{{1\over 4}(\hat{\Phi}- {\Phi})}\; 
\hat{E}^{ {\beta}1}\wedge \hat{E}^{ {\gamma}1} 
\sigma^{  {\#}}_{ {\beta} {\gamma}}] 
- \\ &  
\nonumber -  i e^{{1\over 2}{\Phi}}  
({E}^{ {*}}  - e^{{1\over 4}(\hat{\Phi}- {\Phi})}\; 
\hat{E}^{ {\#}}) 
& \wedge [{E}^{ {\beta}2}\wedge {E}^{ {\gamma}2} 
\sigma^{  {\#}}_{ {\beta} {\gamma}} 
+ e^{{1\over 4}(\hat{\Phi}- {\Phi})} \;
\hat{E}_{ {\beta}}^2\wedge \hat{E}_{ {\gamma}}^2  
\tilde{\sigma}^{ {\#} {\beta} {\gamma}}] =
\\ & & 
\nonumber   {}\qquad {} \qquad {} 
\qquad  = \; {\cal O} (
{E}^{ {a}}\wedge {E}^{ {b}}\; ,\;  
{E}^{ {a}}\wedge \hat{E}^{ {\#}})\; . 
\end{eqnarray}
In Eq. (\ref{hH3II2}) ${\cal O} (
{E}^{ {a}}\wedge {E}^{ {b}}\; ,\;  
{E}^{ {a}}\wedge \hat{E}^{ {\#}})$ denotes the terms 
containing at least two bosonic supervielbein forms 
of 
the {\sl underlying superspace} 
${\cal M}^{(11|32)}$ (\ref{M11}).  
As only $11$ bosonic supervielbein forms, 
{\it e.g.} (\ref{M11EbB}), can be considered as 
independent on ${\cal M}^{(11|32)}$, Eq. (\ref{hH3II2}) implies 
\begin{eqnarray}\label{eqFE1}
{E}^{ {\beta}1}\wedge {E}^{ {\gamma}1} 
\sigma_{  {\tilde{a}} {\beta} {\gamma}} 
 & -   {E}^{ {\beta}2}\wedge {E}^{ {\gamma}2} 
\sigma_{  {\tilde{a}} {\beta} {\gamma}} 
= & \\ \nonumber  
& =  e^{{1\over 4}(\hat{\Phi}- {\Phi})} \; 
\hat{E}^{ {\beta}1}\wedge \hat{E}^{ {\gamma}1} 
\sigma_{  {\tilde{a}} {\beta} {\gamma}} \; 
  & -    e^{{1\over 4}(\hat{\Phi}- {\Phi})} \;
\hat{E}_{ {\beta}}^2\wedge \hat{E}_{ {\gamma}}^2  
\tilde{\sigma}_{ {\tilde{a}}}
^{ {\beta} {\gamma}}  + 
{\cal O} ({E}^{ {a}}\; ,\; \hat{E}^{ {\#}})\; , \qquad  
\\ \label{eqFE2} 
{E}^{ {\beta}1}\wedge {E}^{ {\gamma}1} 
\sigma^{  {\#}}_{ {\beta} {\gamma}} 
 & -   e^{{1\over 4}(\hat{\Phi}- {\Phi})}\; 
\hat{E}^{ {\beta}1}\wedge \hat{E}^{ {\gamma}1} 
\sigma^{  {\#}}_{ {\beta} {\gamma}}
 & =  {\cal O} (
{E}^{ {a}}\; ,\;  \hat{E}^{ {\#}})
\; , \qquad  
\\ \label{eqFE3} 
 \; {E}^{ {\beta}2}\wedge {E}^{ {\gamma}2} 
\sigma^{  {\#}}_{ {\beta} {\gamma}} 
& +  \;  e^{{1\over 4}(\hat{\Phi}- {\Phi})} \;
\hat{E}_{ {\beta}}^2\wedge \hat{E}_{ {\gamma}}^2  
\tilde{\sigma}^{ {\#} {\beta} {\gamma}} 
 & = {\cal O} (
{E}^{ {a}}\; ,\;  \hat{E}^{ {\#}})\; . \qquad 
\end{eqnarray}

Hence Eqs. (\ref{eqFE1})--(\ref{eqFE3}) suggest the 
following relation between the type IIA and type IIB 
fermionic supervielbein forms  
\begin{eqnarray}\label{Td-EhE10}
e^{{1\over 8}{\Phi}}
{E}^{ {\beta}1[-]} = \;  e^{{1\over 8}\hat{\Phi}}
\; 
(\hat{E}^{{\beta}1[-]} + 
\hat{E}^{ {\tilde{a}}(-)} 
\nu^{ {\beta}1}_{ {\tilde{a}}}) 
\; , \qquad \\  
\label{Td-EhE20}
e^{{1\over 8}{\Phi}} {E}^{ {\beta}2[-]}= \; 
e^{{1\over 8}\hat{\Phi}}
\;  
\tilde{\sigma}^{ {\#} {\beta} {\gamma}}
\; ( \hat{E}_{{\gamma}}^{2[-]}  + 
\hat{E}^{ {\tilde{a}}(-)}
\nu_{ {\tilde{a}}}{}^2_{ {\gamma}})
\; ,  
\end{eqnarray}
where $\nu^{ {\beta}1}_{ {\tilde{a}}}$ and 
$\nu_{ {\tilde{a}}}{}^2_{ {\gamma}}$ are 
indefinite coefficients. 
Below we will find their 
explicit form from the torsion constraints 
(\ref{TaIIA=}) and (\ref{TaIIB=}).

In conclusion of this section we note that 
Eqs. (\ref{Td-EhE10}), (\ref{Td-EhE20}) are sufficient to 
find the relation between 
covariant spinor derivatives 
acting on a scalar superfield $V(\tilde{Z})$ defined on the 
nine--dimensional superspace ${\cal M}^{(9|32)}$ (\ref{M9}). 
The differential acting on such a superfield  $V(\tilde{Z})$,  
$dV(\tilde{Z}) = d\tilde{Z}^{\tilde{M}}\partial_{\tilde{M}}V(\tilde{Z})$, 
can be decomposed either on  
type IIA or  on type IIB supervielbein forms, 
\begin{eqnarray} 
\label{dV=AB}
dV(\tilde{Z}) = d\tilde{Z}^{\tilde{M}}\partial_{\tilde{M}}V(\tilde{Z}) 
& = \hat{E}^{ {\tilde{a}}(-)} 
\hat{\nabla}_{ {\tilde{a}}}V(\tilde{Z})+ 
\hat{E}^{ {\alpha}1[-]}
\hat{\nabla}_{{\alpha}1}V(\tilde{Z})+ 
\hat{E}^{2[-]}_{\alpha}
\hat{\nabla}_2^{{\alpha}}V(\tilde{Z})= 
\nonumber \\ 
& =
E^{{\tilde{a}}(-)} 
{\nabla}_{ {\tilde{a}}}V(\tilde{Z})+ 
{E}^{ {\alpha}1[-]} 
{\nabla}_{{\alpha}1}V(\tilde{Z}) + 
{E}^{ {\alpha}2[-]} 
{\nabla}_{{\alpha}2}V(\tilde{Z})
\; . \qquad 
\end{eqnarray}
Substituting the T--duality rules (\ref{T-EIIII}), 
(\ref{Td-EhE10}), (\ref{Td-EhE20}), 
one finds that 
Eq. (\ref{dV=AB}) implies, in particular, 
\begin{eqnarray} 
\label{dfV=AB}
& e^{- {1 \over 8}\Phi} {\nabla}_{{\alpha}1}V(\tilde{Z}) = 
e^{- {1 \over 8}\hat{\Phi}} \hat{\nabla}_{{\alpha}1}V(\tilde{Z})\; , 
\quad 
e^{- {1 \over 8}\Phi}  
{\nabla}_{{\alpha}2}V(\tilde{Z}) = - e^{- {1 \over 8}\hat{\Phi}} 
\sigma^{ {\#}}_{ {\alpha} {\beta}}
\hat{\nabla}_2^{ {\beta}}V(\tilde{Z})\; . 
\end{eqnarray}

\subsection{T--duality rules for fermionic forms from 
the supergravity constraints II. 
Complete form from torsion constraints}

First, let us observe that the T--duality rule 
(\ref{T-EIIII}) implies the following relation between 
the type IIA and type IIB representations for the torsion of 
nine--dimensional superspace ${\cal M}^{(9|32)}$ (\ref{M9}) 
(see {\it l.h.s}'s of Eqs. 
(\ref{Ta9IIA}) and (\ref{Ta9IIB}))
\begin{eqnarray} 
\label{T-Ta9} 
e^{{\Phi \over 4}}{\cal D}^{[-]} {E}^{\tilde{a}(-)}
= 
e^{{\hat{\Phi} \over 4}} [ \hat{{\cal D}}^{[-]} \hat{E}^{\tilde{a}(-)} 
+ {1\over 4} \hat{E}^{\tilde{a}(-)}\wedge d(\hat{\Phi}-{\Phi}) + 
 \hat{E}^{\tilde{b}(-)}\wedge (\hat{w}^{[-]}{}_{\tilde{b}}{}^{\tilde{a}}- 
{w}^{[-]}{}_{\tilde{b}}{}^{\tilde{a}})]\; .
\end{eqnarray}
Using the constraints  (\ref{Ta9IIA}) and (\ref{Ta9IIB}) 
and the relations between fermionic forms 
(\ref{Td-EhE10}), (\ref{Td-EhE20}) one finds that the lower dimensional 
components of the Eq. (\ref{T-Ta9}) are satisfied identically, while the 
dimension $3/2$ components provide one with the equations for 
 $\nu^{ {\beta}1}_{ {\tilde{a}}}$ and 
$\nu_{ {\tilde{a}}}{}^2_{ {\gamma}}$,  
\begin{eqnarray}\label{Eqnu-1}
2i \tilde{\sigma}^{\tilde{a}}_{\alpha\beta} 
\nu^{ {\beta}1}_{ {\tilde{a}}} = {1\over 4} \delta_{\tilde{b}}{}^{\tilde{a}}
(\hat{\nabla}_{\alpha 1}\hat{\Phi}- e^{{1\over 8} (\hat{\Phi}- \Phi)}
{\nabla}_{\alpha 1}{\Phi}) 
+ \Delta \hat{w}^{[-]}{}_{\alpha 1 \, \tilde{b}}{}^{\tilde{a}}\; , 
\qquad \\  
\label{Eqnu2} 
2i \tilde{\sigma}^{\tilde{a}\alpha\beta} 
\nu_{ {\tilde{b}}}{}^2_{ {\gamma}}
= 
{1\over 4} \delta_{\tilde{b}}{}^{\tilde{a}}
(\hat{\nabla}^{\alpha}_2 \hat{\Phi}- e^{{1\over 8} (\hat{\Phi}- \Phi)}
\tilde{\sigma}^{\# \alpha\beta}
{\nabla}_{\beta 2}{\Phi}) 
+ \Delta \hat{w}^{[-]}{}^{\alpha}_2{}_{\tilde{b}}{}^{\tilde{a}}\; ,
\end{eqnarray}
where 
$\Delta \hat{w}^{[-]}{}_{\tilde{b}}{}^{\tilde{a}}\equiv
(\hat{w}^{[-]}- {w}^{[-]}){}_{\tilde{b}}{}^{\tilde{a}}= 
\hat{E}^{\tilde{c}(-)}\Delta \hat{w}^{[-]}{}_{\tilde{c}\tilde{b}}{}^{\tilde{a}}
+ \hat{E}^{\alpha 1 [-]} 
\Delta \hat{w}^{[-]}{}_{\alpha 1 \, \tilde{b}}{}^{\tilde{a}}
+ \hat{E}^{2[-]}_{\alpha}
\Delta \hat{w}^{[-]}{}^{\alpha}_2{}_{\tilde{b}}{}^{\tilde{a}}$. 
The solutions of these equations, 
\begin{eqnarray}\label{Snu1}
\nu^{ {\beta}1}_{ {\tilde{a}}} = - {i\over 8} 
\tilde{\sigma}_{\tilde{a}}^{\alpha\beta} 
(\hat{\nabla}_{\alpha 1}\hat{\Phi}- e^{{1\over 8} (\hat{\Phi}- \Phi)}
{\nabla}_{\alpha 1}{\Phi}) \; , \qquad 
\\  
\label{Snu2} 
\nu_{ {\tilde{a}}}{}^2_{ {\alpha}} = - {i\over 8}  
{\sigma}_{\tilde{a}\alpha\beta} 
(\hat{\nabla}^{\alpha}_2 \hat{\Phi}- e^{{1\over 8} (\hat{\Phi}- \Phi)}
\tilde{\sigma}^{\# \alpha\beta} {\nabla}_{\beta 2}{\Phi}) \; , \qquad
\end{eqnarray} 
provide us with 
the following  final form of the essential fermionic 
T--duality rules  
\begin{eqnarray}\label{Td-EhE1}
e^{{1\over 8}{\Phi}} 
({E}^{ {\beta}1[-]} - {i \over 8} 
{E}^{ {\tilde{a}}(-)} 
\tilde{\sigma}_{ {\tilde{a}}}{}^{ {\beta} {\gamma}}
{\nabla}_{ {\gamma}1}{\Phi}) 
= \;  e^{{1\over 8}\hat{\Phi}}\; 
(\hat{E}^{{\beta}1[-]} - {i \over 8} \hat{E}^{ {\tilde{a}}(-)}
\tilde{\sigma}_{ {\tilde{a}}}{}^{ {\beta} {\gamma}}
\hat{\nabla}_{ {\gamma}1}\hat{\Phi}) 
\; , \qquad \\  
\label{Td-EhE2} 
e^{{1\over 8}{\Phi}} 
({E}^{ {\beta}2[-]}
- {i \over 8} 
{E}^{ {\tilde{a}}(-)} 
\tilde{\sigma}_{ {\tilde{a}}}{}^{ {\beta} {\gamma}}
{\nabla}_{ {\gamma}2}{\Phi})
= \; e^{{1\over 8}\hat{\Phi}}\;  
\tilde{\sigma}^{ {\#} {\beta} {\gamma}}
\; ( \hat{E}_{ {\gamma}}^{2[-]}  - {i \over 8}  
\hat{E}^{{\tilde{a}}(-)}
{\sigma}_{\tilde{ {a}} {\beta} {\gamma}}
\hat{\nabla}^{ {\gamma}}_2\hat{\Phi}) \; . 
\end{eqnarray}

Note that if we used (\ref{dV=AB}) to decompose $d\Phi$ in 
Eq. (\ref{T-Ta9}) on the type IIA superforms, 
Eqs. (\ref{Snu1}), (\ref{Snu2}) 
would read 
\begin{eqnarray}\label{Snu1-}
\nu^{ {\beta}1}_{ {\tilde{a}}} = - {i\over 8} 
\tilde{\sigma}_{\tilde{a}}^{\alpha\beta} 
\hat{\nabla}_{\alpha 1}(\hat{\Phi}- {\Phi}) \; , \qquad 
\\  
\label{Snu2-} 
\nu_{ {\tilde{a}}}{}^2_{ {\alpha}} = - {i\over 8}  
{\sigma}_{\tilde{a}\alpha\beta} 
\hat{\nabla}^{\alpha}_2 (\hat{\Phi}- {\Phi}) \; , \qquad
\end{eqnarray} 
where $(\hat{\Phi}- {\Phi})$ could be expressed through the type IIA 
superfields by using the T--duality rule (\ref{T-Phi}), 
\begin{eqnarray}\label{hP-P}
\hat{\Phi}(\tilde{Z})- {\Phi}(\tilde{Z})= ln(e^{{1\over 4}\hat{\Phi}}
i_{\hat{z}}\hat{E}^{\#})\; .
\end{eqnarray} 
Such notation allows us
to rewrite the rules (\ref{Td-EhE1}), (\ref{Td-EhE2})    
in slightly different form, 
see Eqs. (\ref{Td-EhE1-}), (\ref{Td-EhE2-}) below.  
As a by--product, on these stages one also obtains the T--duality rules for 
the `nine--dimensional' part of the spin connections (see (\ref{wabIIA-}), 
(\ref{wabIIB-}))
\begin{eqnarray}\label{Td-w-w}
w^{\tilde{b}\tilde{a} [-]}  = 
\hat{w}^{\tilde{b}\tilde{a} [-]}+ 
{1\over 4} \hat{E}^{\alpha 1[-]} 
(\sigma^{\tilde{b}\tilde{a}})_{\alpha }{}^{\beta} \hat{\nabla}_{\beta 1} 
(\hat{\Phi}- {\Phi}) - 
{1\over 4} \hat{E}_{\alpha }^{2[-]} 
(\sigma^{\tilde{b}\tilde{a}})_{\beta}{}^{\alpha } \hat{\nabla}^{\beta}_2
(\hat{\Phi}- {\Phi}) + \nonumber \\  + 
{E}^{[\tilde{b}(-)} \hat{\nabla}^{\tilde{a}]} 
(\hat{\Phi}- {\Phi}) 
- {3i\over 64} {E}_{\tilde{c}}^{(-)} 
\, \left(\tilde{\sigma}^{\tilde{c}\tilde{b}\tilde{a}}{}^{\alpha\beta} 
\hat{\nabla}_{\alpha 1} (\hat{\Phi}- {\Phi})
\, \hat{\nabla}_{\beta 1} (\hat{\Phi}- {\Phi}) +\right. 
\nonumber \\  \qquad + \left.
{\sigma}^{\tilde{c}\tilde{b}\tilde{a}}{}_{\alpha\beta} 
\hat{\nabla}^{\alpha}_2 (\hat{\Phi}- {\Phi})\, \hat{\nabla}^{\beta }_2 
(\hat{\Phi}- {\Phi}) \right) 
\; , 
\end{eqnarray}
where one can substitute (\ref{hP-P})
for $(\hat{\Phi}- {\Phi})$.

To complete the fermionic T--duality rules we have to find 
the relation between the fermionic superfields 
$i_{\hat{z}}\hat{E}^{\alpha 1}$, $i_{\hat{z}}\hat{E}_{\alpha}^2$ 
and  $i_yE^{\alpha 1,2}$. As that are composed, 
Eqs.  (\ref{chiIIA}), (\ref{chiIIB}), 
to this end it is sufficient to use the relation 
between covariant spinor derivatives, 
Eqs. (\ref{dfV=AB}),  for $V(\tilde{Z})= 
\hat{E}^{\#}_{\hat{z}}(\tilde{Z}) \equiv i_{\hat{z}}\hat{E}^{\#}$ and 
$V(\tilde{Z})= \hat{E}_y^{*}(\tilde{Z})\equiv i_{y}\hat{E}^{*}$.
In such a way one finds 
the following T--duality rules 
for the fermionic superfields $E_{\hat{z}}^{{{\a}1}}(\tilde{Z})\equiv 
i_{\hat{z}}\hat{E}^{{{\a}1}}$, $
\hat{E}_{\hat{z}}{}^2_{{{\a}}}(\tilde{Z})\equiv 
i_{\hat{z}}\hat{E}^2_{ {{\a}}}$ and ${E}_y^{ {{\a}}1,2}(\tilde{Z})\equiv 
i_y{E}^{ {{\a}}1,2}$:
\begin{eqnarray}\label{Td-ch1}
e^{-{1\over 8}{\Phi}} 
\left({i_{y}{E}^{{\beta}1} \over i_yE^*}
+ {i \over 8} 
\tilde{\sigma}^{ {*}}{}^{ {\beta} {\gamma}}
{\nabla}_{ {\gamma}1}{\Phi}\right) 
= - \,  e^{-{1\over 8}\hat{\Phi}}\; 
\left(
{i_{\hat{z}}\hat{E}^{ {{\b}1}}\over i_{\hat{z}}\hat{E}^{ {\#}}} 
+ {i \over 8} 
\tilde{\sigma}^{ {\#}}{}^{ {\beta} {\gamma}}
\hat{\nabla}_{ {\gamma}1}\hat{\Phi}\right) 
\; , \qquad \\  
\label{Td-ch2}
e^{-{1\over 8}{\Phi}} 
\left({i_y{E}^{ {\beta}2}\over i_yE^*} + {i \over 8} 
\tilde{\sigma}^{ {*} {\beta} {\gamma}}
{\nabla}_{ {\gamma}2}{\Phi}\right) = 
\; e^{- {1\over 8}\hat{\Phi}}\;  
\tilde{\sigma}^{ {\#} {\beta} {\gamma}}
\; \left( {i_{\hat{z}}\hat{E}_{ {\gamma}}^2 \over i_{\hat{z}}\hat{E}^{ {\#}}} 
+ {i \over 8}  
{\sigma}^{\#}_{ {\beta} {\gamma}}
\hat{\nabla}^{ {\gamma}}_2\hat{\Phi}\right) \; . 
\end{eqnarray}

Eqs. (\ref{Td-EhE1}), (\ref{Td-EhE2})  
can be collected together with 
Eqs. (\ref{Td-ch1}), (\ref{Td-ch2}) in the following 
fermionic T-duality rules involving only the {\sl complete} forms 
and their contractions
\begin{eqnarray}\label{Td-EhE1!}
e^{{1\over 8}{\Phi}} 
({E}^{ {\beta}1} - {i \over 8} 
{E}^{ {a}} 
\tilde{\sigma}_{ {a}}{}^{ {\beta} {\gamma}}
{\nabla}_{ {\gamma}1}{\Phi}) 
= \;  e^{{1\over 8}\hat{\Phi}}\; 
(\hat{E}^{ {\beta}1} - {i \over 8} 
\hat{E}^{ {a}}
\tilde{\sigma}_{ {a}}{}^{ {\beta} {\gamma}}
\hat{\nabla}_{ {\gamma}1}\hat{\Phi}) - 
\;  \qquad \nonumber \\  
{} \qquad - 
 \;  e^{{1\over 8}\hat{\Phi}}\; 
\left(\hat{E}^{ {\#}} + 
e^{{1\over 4}(\Phi -\hat{\Phi})}
{E}^{ {*}}\right) \; \left(
{i_{\hat{z}}
\hat{E}^{ {\beta}1}\over i_{\hat{z}}\hat{E}^{ {\#}}}
+  {i \over 8} 
\tilde{\sigma}^{ {\#}}{}^{ {\beta} {\gamma}}
\hat{\nabla}_{ {\gamma}1}\hat{\Phi}\right) \; , 
\\ 
\label{Td-EhE2!} 
e^{{1\over 8}{\Phi}} 
({E}^{ {\beta}2} - {i \over 8} 
{E}^{ {a}} 
\tilde{\sigma}_{ {a}}{}^{ {\beta} {\gamma}}
{\nabla}_{ {\gamma}2}{\Phi}) 
= \; 
e^{{1\over 8}\hat{\Phi}}\;  
\tilde{\sigma}^{ {\#} {\beta} {\gamma}}
\; ( \hat{E}_{ {\gamma}}^2  - {i \over 8}  
\hat{E}^{ {a}}
{\sigma}_{{ {a}} {\beta} {\gamma}}
\hat{\nabla}^{ {\gamma}}_2\hat{\Phi}) - 
 \qquad \nonumber \\  
{} \qquad -   e^{{1\over 8}\hat{\Phi}}\,  
\left(\hat{E}^{ {\#}} - 
e^{{1\over 4}(\Phi -\hat{\Phi})}
{E}^{ {*}}\right) \, 
\tilde{\sigma}^{ {\#} {\beta} {\gamma}}
\left({i_{\hat{z}}\hat{E}_{ {\gamma}}^2
\over i_{\hat{z}}\hat{E}^{ {\#}}}  + {i \over 8}  
{\sigma}^{\#}_{ {\beta} {\gamma}}
\hat{\nabla}^{ {\gamma}}_2\hat{\Phi}\right) \; . 
\end{eqnarray}
Indeed, due to the last terms in (\ref{Td-EhE1!}), (\ref{Td-EhE2!}), 
the relations obtained by 
contractions of these equations with $i_{\hat{z}}$ are 
satisfied identically, while the contractions 
of these equations with $i_{y}$ reproduce the T-duality rules 
for spinor superfields  (\ref{Td-ch1}), (\ref{Td-ch2}). 
The parts of Eqs. (\ref{Td-EhE1!}), (\ref{Td-EhE2!}) which do not 
contain neither $\hat{E}^{ {\#}}$ nor ${E}^{ {*}}$
reproduce Eqs. (\ref{Td-EhE1}), (\ref{Td-EhE2}).

It might be  useful to rewrite the fermionic T--duality rules 
in the more standard form similar to the one of Eqs. 
(\ref{T-EIIII})--(\ref{T-B-}). 
To this end one uses (\ref{dfV=AB})  and (\ref{T-Phi}) to 
present Eqs. (\ref{Td-ch1}), (\ref{Td-ch2})  and 
(\ref{Td-EhE1}), (\ref{Td-EhE2}) as 
\begin{eqnarray}\label{Td-ch1-}
e^{-{1\over 8}{\Phi}} 
{i_{y}{E}^{{\beta}1} \over i_yE^*} 
&=& - \,  e^{-{1\over 8}\hat{\Phi}}\; 
\left(
{i_{\hat{z}}\hat{E}^{ {{\b}1}}\over i_{\hat{z}}\hat{E}^{ {\#}}} 
+ {i \over 4} 
\tilde{\sigma}^{ {\#}}{}^{ {\beta} {\gamma}}
\hat{\nabla}_{ {\gamma}1}\hat{\Phi}- {i \over 8} 
\tilde{\sigma}^{ {\#}}{}^{ {\beta} {\gamma}}
\hat{\nabla}_{ {\gamma}1} ln \left(e^{\hat{\Phi}\over 4} i_{\hat{z}}E^{\#}
\right)\right) \; , \qquad \\  
\label{Td-ch2-}
e^{-{1\over 8}{\Phi}} 
{i_y{E}^{ {\beta}2}\over i_yE^*} &=&
\; e^{- {1\over 8}\hat{\Phi}}\;  
\tilde{\sigma}^{ {\#} {\beta} {\gamma}}
\; \left( {i_{\hat{z}}\hat{E}_{ {\gamma}}^2 \over i_{\hat{z}}\hat{E}^{ {\#}}} 
+ {i \over 4}  
{\sigma}^{\#}_{ {\beta} {\gamma}}
\hat{\nabla}^{ {\gamma}}_2\hat{\Phi} - 
{i \over 8}  
{\sigma}^{\#}_{ {\beta} {\gamma}}
\hat{\nabla}^{ {\gamma}}_2
ln \left({e^{\hat{\Phi}\over 4}} i_{\hat{z}}E^{\#}\right)
\right) \;   \qquad 
\end{eqnarray}
and
\begin{eqnarray}\label{Td-EhE1-}
e^{{1\over 8}{\Phi}} 
{E}^{ {\beta}1[-]} 
&=& \;  e^{{1\over 8}\hat{\Phi}}\; 
\left(\hat{E}^{{\beta}1[-]} - {i \over 8} \hat{E}^{ {\tilde{a}}(-)}
\tilde{\sigma}_{ {\tilde{a}}}{}^{ {\beta} {\gamma}}
\hat{\nabla}_{ {\gamma}1} ln \left({e^{\hat{\Phi}\over 4}} i_{\hat{z}}E^{\#}
\right)
\right)
\; , \qquad \\  
\label{Td-EhE2-} 
e^{{1\over 8}{\Phi}} 
{E}^{ {\beta}2[-]}
&=& \; e^{{1\over 8}\hat{\Phi}}\;  
\tilde{\sigma}^{ {\#} {\beta} {\gamma}}
\; \left( \hat{E}_{ {\gamma}}^{2[-]}  - {i \over 8}  
\hat{E}^{{\tilde{a}}(-)}
{\sigma}_{\tilde{ {a}} {\beta} {\gamma}}
\hat{\nabla}^{ {\gamma}}_2
ln \left({e^{\hat{\Phi}\over 4}} i_{\hat{z}}E^{\#}\right)\right)
\; . \qquad 
\end{eqnarray}

Let us comment on the relation of the above results with the 
T--duality rules for fermionic superforms 
{\sl in the string frame} (see Eq. (\ref{Estr})) derived in Ref. \cite{kulik}. 
To this end, at first, one ignores the  dilaton superfield in 
Eqs. (\ref{Td-ch1}), (\ref{Td-ch2}) 
and, at second, one passes to the general  supervielbein forms 
(not adapted to the isometries). The result is 
\begin{eqnarray}\label{T-strf9}
{\cal E}_y^{\beta 1} = {\hat{{\cal E}}_{\hat{z}}^{{\beta}1}\over 
\hat{{\cal G}}_{\hat{z}\hat{z}}}\; , \qquad 
{\cal E}_y^{\beta 2} =  
{\hat{{\cal E}}_{\hat{z}}^{{a}}\tilde{\sigma}_a^{\beta\gamma}
\hat{{\cal E}}_{\hat{z}}{}_{\gamma}^{2}\over 
\sqrt{|\hat{{\cal G}}_{\hat{z}\hat{z}}|}
\hat{{\cal G}}_{\hat{z}\hat{z}}}\; , \quad 
\end{eqnarray} 
where 
\begin{eqnarray}\label{calG}
\hat{{\cal G}}_{\hat{z}\hat{z}}\equiv \hat{{\cal E}}_{\hat{z}}^{\hat{a}}
\hat{{\cal E}}_{\hat{z}\hat{a}}\; .
\end{eqnarray}
In the same manner, ignoring the inputs from dilaton superfields in 
Eqs. (\ref{Td-EhE1}), (\ref{Td-EhE2}), and passing form $E^{[-]}$ to 
$E^{(-)}$ by the use of (\ref{EfIIA-}), (\ref{EfIIB-}), we arrive at 
\begin{eqnarray}\label{T-strf-}
{\cal E}^{\beta 1(-)} &=& \hat{{\cal E}}^{\beta 1(-)}
-  {\hat{{\cal E}}^{a(-)} i_{\hat{z}}\hat{{\cal E}}_a 
\over \hat{{\cal G}}_{\hat{z}\hat{z}} }
\, i_{\hat{z}} \hat{{\cal E}}^{{\beta}1} +
{i_{\hat{z}} \hat{B}_2 \over \hat{{\cal G}}_{\hat{z}\hat{z}}}
\, i_{\hat{z}} \hat{{\cal E}}^{{\beta}1}
\; , \quad \nonumber \\ 
{\cal E}^{\beta 2(-)} &=& -  { i_{\hat{z}} \hat{{\cal E}}^{\hat{a}} 
\tilde{\sigma}_{\hat{a}}^{\beta\gamma} \over 
\sqrt{|\hat{{\cal G}}_{\hat{z}\hat{z}}|}}
\left(\hat{{\cal E}}_{\gamma}^{2(-)} - 
{\hat{{\cal E}}^{a(-)} i_{\hat{z}}\hat{{\cal E}}_a 
\over \hat{{\cal G}}_{\hat{z}\hat{z}} } \, 
i_{\hat{z}}\hat{{\cal E}}_{\gamma}^{2}
- {i_{\hat{z}} \hat{B}_2 \over \hat{{\cal G}}_{\hat{z}\hat{z}}} 
\, i_{\hat{z}}\hat{{\cal E}}_{\gamma}^{2}\right) . \quad 
\end{eqnarray} 
Eqs. (\ref{T-strf9}), (\ref{T-strf-}) coincide with the T--duality rules 
for the fermionic supervielbein forms presented in Ref. 
\cite{kulik}\footnote{An evident factor  
$1/\sqrt{|\hat{{\cal G}}_{\hat{z}\hat{z}}|}$ 
in Eqs. (\ref{T-strf9}), (\ref{T-strf-}) for ${\cal E}^{\beta 2}$ 
should be restored in Eqs. of Ref. \cite{kulik}.}.

\section{Consistency of the T--duality rules with the constraints for 
RR field strengths} 
\setcounter{equation}0

Thus all the T-duality rules are restored with the use 
of the torsion constraints and the constraints for the field strengths 
of the NS--NS gauge superforms. The question remains: whether these  
 the T--duality rules, including 
Eqs. (\ref{Td-EhE1!}), (\ref{Td-EhE2!}) and (\ref{rC=}), 
 are consistent with the constraints 
for the field strengths 
\begin{equation}\label{R=R+R+}
R=dC - C \wedge H_3 = 
R_1 \oplus R_3 \oplus R_5  \oplus R_7  \oplus R_9  \; , 
\end{equation}
\begin{equation}\label{hR=hR+hR+}
\hat{R}=d\hat{C} - \hat{C} \wedge \hat{H}_3 = 
\hat{R}_2 \oplus \hat{R}_4 \oplus \hat{R}_6  \oplus \hat{R}_8  \oplus 
\hat{R}_{10} \; ,  
\end{equation}
of the RR superforms (\ref{rC=C+C+}), (\ref{rCIIB}) and 
(\ref{rhC=hC+hC+}), (\ref{rCIIA}). 
These constraints can be found in \cite{c0}. 
For our consideration it is convenient to collect them 
in the following equations for the formal sums of the 
RR field strengths
\begin{equation}\label{hR=hc}
\hat{R}=d\hat{C} - \hat{C} \wedge \hat{H}_3 = 
 2i e^{-{3\over 4}\hat{\Phi}} \; 
\hat{E}^{ {\a}1}
\wedge
\hat{E}^2_{ {\b}}
\wedge \hat{\bar\gamma}(\hat{\Phi})^{ {\b}}{}_{ {\a}}
  + \ldots \;   ,
\end{equation}
\begin{equation}\label{R=c} 
R=dC - C \wedge H_3 = 
2i e^{-{3\over 4}{\Phi}} \; 
{E}^{ {\a}2}
\wedge
{E}^{ {\b}1}
\wedge \bar{\sigma}({\Phi})_{ {\a} {\b}}
   + \ldots \;   .  
\end{equation}
Here the matrix valued formal sums of differential (super)forms 
$\hat{\bar\gamma}(\hat{\Phi}){}^{ {\b}}{}_{ {\a}}$, 
$\bar{\sigma}({\Phi})_{ {\a} {\b}}$ are defined by 
\begin{eqnarray}\label{hg=sum}
& \hat{\bar\gamma}(\hat{\Phi}){}^{ {\b}}{}_{ {\a}}
\equiv \delta^{ {\b}} {}_{ {\a}}
\oplus \hat{\bar\gamma}^{(2)}(\hat{\Phi})^{ {\b}}{}_{ {\a}}
\oplus \hat{\bar\gamma}^{(4)}(\hat{\Phi})^{ {\b}}{}_{ {\a}}
\oplus \hat{\bar\gamma}^{(6)}(\hat{\Phi})^{ {\b}}{}_{ {\a}}
\oplus \hat{\bar\gamma}^{(8)}(\hat{\Phi})^{ {\b}}{}_{ {\a}}
\\ & \label{hg2n}
\hat{\bar\gamma}^{(2n)}(\hat{\Phi})^{ {\b}}{}_{ {\a}}
= {e^{{n\over 2}\hat{\Phi} }\over (2n)!} 
\hat{E}^{ {a}_{2n}}
\wedge \ldots \wedge \hat{E}^{ {a}_{1}}
\sigma_{ {a}_{1} \ldots {a}_{2n}}
{}^{ {\b}}{}_{ {\a}}\; , 
\end{eqnarray}
\begin{eqnarray}\label{bs=sum}
& \bar{\sigma}({\Phi})_{ {\a} {\b}} 
= 
\bar{\sigma}({\Phi})^{(1)}_{ {\a} {\b}}  \oplus 
\bar{\sigma}({\Phi})^{(3)}_{ {\a} {\b}}  \oplus 
\bar{\sigma}({\Phi})^{(5)}_{ {\a} {\b}}  \oplus 
\bar{\sigma}({\Phi})^{(7)}_{ {\a} {\b}}  \; , \\ 
& \label{bs2n1}
\bar{\sigma}({\Phi})^{(2n+1)}_{ {\a} {\b}} 
\equiv 
 {e^{{2n+1\over 4}{\Phi} }\over (2n+1)!} 
{E}^{ {\hat{a}}_{2n+1}}
\wedge \ldots \wedge {E}^{ {\hat{a}}_{1}}
\sigma_{ {\hat{a}}_{1} \ldots {\hat{a}}_{2n+1} {\a}
 {\b}}\; . 
\end{eqnarray}
The terms denoted by ellipses in Eqs. (\ref{hR=hc}), (\ref{R=c})
include {\sl not more than one}  fermionic supervielbein form
and can be ignored due to the following reasons.  
When the expressions for certain differential q--forms ($q>2$) are extracted 
from  (\ref{hR=hc}), (\ref{R=c}), the terms with less than two 
fermionic supervielbein forms contain more bosonic supervielbein forms  
and, thus, describe  the higher dimensional components of the 
q--form equation. 
However, as in the case of NS--NS superfield strengths (see Sec. 5.1), 
such higher dimensional components  
can be derived as a consequence of the 
lowest dimensional ones and the torsion constraints 
(\ref{TaIIA=}), (\ref{TaIIB=})  
with the use of superspace Bianchi identities 
(which can be collected in $d\hat{R}\equiv \hat{R}\wedge \hat{H}_3$ 
and $d{R}\equiv {R}\wedge {H}_3$). 
Thus we can conventionally ignore them in the analysis of T--duality as there 
consistency is guarantied provided the lowest dimensional equations 
are consistent with the T--duality rules.  
This is a `bottom--up' form of the spinor cohomology approach of Sec. 5.1
\footnote{The constraint 
$R_1 =  dC_0 =
e^{- \Phi}
 E^{\underline{\a}1}
\nabla_{\underline{\a}2}\Phi
-
e^{- \Phi}
E^{\underline{\a}2}
\nabla_{\underline{\a}1}\Phi
+
E^{\underline{a}} R_{\underline{a}}$ for the axion `field strength', 
which is completely hidden in ellipses in Eq. (\ref{R=c}),  
also can be derived from the torsion constraints.}.

Having in hands the explicit T--duality rules for the RR fields, 
Eq. \p{rC=}, one can 
obtain by direct calculations the T--duality rules for their field strengths.
To this end one can  
\begin{itemize}
\item 
take the derivative of Eq. (\ref{rC=});  
\item use  the conditions of isometry for 
NS--NS two forms and RR forms 
$$ d(i_y B_2)= - i_y H_3\; , \quad 
d(i_{\hat{z}} \hat{B}_2)= - i_{\hat{z}} \hat{H}_3\; , \quad 
d(i_{y} {C})= - i_{y} (d{C})\; , \quad  
d(i_{\hat{z}} \hat{C})= - i_{\hat{z}} (d\hat{C})\; , $$ 
to arrive at the expression in terms of the field strength; 
\item use Eqs. 
(\ref{R=c}), (\ref{hR=hc}) to obtain the expressions in terms 
of generalized field strength 
$R$ and  $\hat{R}$ instead of $dC$, $d\hat{C}$; 
\item observe that in the result of such calculations
all the terms involving potential(s) $C$ (after the use 
of Eq. (\ref{rC=})) 
can be collected in the expression 
$$ C \wedge \left(\hat{H}_3 - H_3 + 
(dy+ E^{ {*}(-)}/i_y E^{ {*}}) 
\wedge i_y H_3- (d\hat{z} + 
\hat{E}^{ {\#}(-)}/ i_{\hat{z}}\hat{E}^{ {\#}}) 
\wedge i_{\hat{z}}\hat{H}_3\right) $$ 
which vanishes in accordance with (\ref{H3=hH3}). 
\end{itemize} 
In such a way we arrive at the T--duality rules for the RR field strengths 
(\ref{R=c}), (\ref{hR=hc}),  
\begin{eqnarray}
  \label{R=hR} 
R =  -  i_{\hat{z}} \hat{R}   +  \left(dy +  
{E^{ {*}(-)}\over i_y E^{ {*}}}\right)\wedge 
\left(\hat{R} + (d\hat{z} + 
{\hat{E}^{ {\#}(-)}\over i_{\hat{z}}\hat{E}^{ {\#}}})
\wedge i_{\hat{z}} \hat{R} \right) \; ,
\end{eqnarray}
which are gauge invariant and resemble the rules 
(\ref{rC=}) for the RR superfield potentials.

Now one can verify that the derived T--duality rules are completely 
consistent  with lower dimensional spin--tensor relations involved into the 
differential form constraints \p{hR=hc}, \p{R=c} for RR field strength. 
To this end one i) substitutes the constraints  
\p{hR=hc}, \p{R=c} into 
\p{R=hR} ii) checks that the resulting equation is satisfied  identically
when the T--duality rules for NS--NS superfields, Eq. 
\p{T-EIIII}--\p{T-B-},  and for the 
fermionic forms, Eqs.  
(\ref{Td-EhE1!}), (\ref{Td-EhE2!}), 
are taken into account. 
In the light of above consideration (on the `bottom--up' form of 
the spinor cohomology approach), on this way one can 
ignore the terms denoted by ellipses in (\ref{hR=hc}) and (\ref{R=c}), 
as well as the terms with less than two fermionic forms which appear 
after substitution of the fermionic T--duality rules into Eqs.   
\p{hR=hc}, \p{R=c}. With this shortcut the explicit check of the consistency 
reduces to a simple exercise in sigma--matrix algebra, which we leave 
for reader.

\section{Conclusion}
\setcounter{equation}0

In this paper we have obtained the complete set of the  
{\sl superfield T--duality rules} which are summarized in the Appendix A 
(see Secs. A1 and A2). 
These are  the relations between all 
superfield potentials of type IIA and type IIB supergravity,  
including fermionic supervielbein forms and 
all the  Ramond--Ramond superfield potentials.
For their derivation we used the 
supervielbeins in the Einstein frame which are 
adapted to the isometries 
(i.e. obey the superfield generalization of the Kaluza--Klein ansatz 
\cite{ansatz}). 
We also present the rules formulated for general  
supervielbeins in the string frame, 
where our results for NS-NS superfields and 
fermionic superforms coincide with the ones obtained 
in \cite{kulik} (Appendix A2). 
Thus we completed the rules from \cite{kulik} by 
the T--duality rules for RR gauge superfields. 
We also propose the differential form representation for 
the T--duality rules (Appendix A3)
which have allowed to verify their consistency 
with the complete set of supergravity constraints. 

Let us stress the  basic observation which has allowed us to 
sort out the problem that hampered the way to the complete superfield   
generalization of the T--duality rules. 
It consists in the possibility to treat the T--duality as a transformation 
of supervielbein and other superforms rather than of the 
superspace coordinates.  
This implies the identification of all the fermionic coordinates of the 
{\sl curved} type IIA and type IIB superspaces, as well as of all but one 
their bosonic coordinates. 
The identification of the fermionic coordinates 
of curved type IIA and type IIB superspaces is possible due to the fact that 
 fermionic coordinates of a curved superspace carry neither spinor 
indices nor chiralities,  but rather are transformed by superspace 
diffeomorphisms. The different chiralities of the fermionic coordinate 
of the {\sl flat}   type IIA and type IIB superspaces originate 
in different chiralities of the fermionic supervielbein forms of 
the {\sl curved} superspaces and can be reproduced in the flat superspace 
limit, after the fermionic supervielbein forms are identified 
with the exterior derivatives of the fermionic coordinates.  
As far as the curved superspaces are concerned, 
assuming that the type IIA and 
type IIB supergravities have one bosonic isometry direction, $\hat{z}$ and $y$ 
respectively, one may consider them as defined on surfaces $y=0$ and 
$\hat{z}=0$ in the underlying eleven--dimensional 
superspace ${\cal M}^{(11|32)}$ (\ref{M11}). 
The intersection of these surfaces gives a nine--dimensional superspace  
${\cal M}^{(9|32)}$ (\ref{M9}). 

Our results  clarify the relation of T--duality with superfield 
formulations of supergravity and, as  we hope, might provide new 
insights in M--theory. 
Our approach  can be also extended to 
the more complicated  $SO(n,n)$ T--duality provided 
the superfield generalization 
of the Kaluza--Klein ansatz for the dimensional reduction 
down to $d=10-n$ dimensions is elaborated for these cases.

\bigskip

{\it Acknowledgments}. 

The authors thank Dima Sorokin and Mario Tonin for reading the manuscript 
and useful suggestions. 
This work has been partially supported by the research grant BFM2002-03681 
from the Ministerio de Ci\'encia y Tecnolog\'{\i}a and
from EU FEDER funds, 
by the Ukrainian FFR grant 
$\# 383$,  the INTAS grant N 2000-254 and the European Community's 
Human Potential Programme under contract HPRN--CT--2000--00131 Quantum 
Spacetime.

\bigskip

\section*{Notice added}
When the present paper has been published, we became award that 
the T--duality rules for the bosonic RR fields 
were obtained for the first type in Ref. \cite{MO98}, 
several months before \cite{simon}.

\newpage 

\section*{Appendix A: Summary of superfield  T--duality rules} 
\renewcommand{\theequation}{A.\arabic{equation}} 
\setcounter{equation}0

\section*{A1. T-duality rules with supervielbein in the Einstein frame 
adapted to the isometry (\ref{EaIIAd})--(\ref{KKE9B})}

For the bosonic supervielbeins adapted to the isometries,  
i.e. obeying Kaluza--Klein ansatz \cite{ansatz}
 \begin{eqnarray}\label{KKA}
& type \; IIA \; : & \hat{E}^{\hat{a}}= (\hat{E}^{\tilde{a}}, \hat{E}^{\#}) 
\; , \qquad 
\hat{E}^{ {\tilde{a}}}= \hat{E}^{ {\tilde{a}}(-)}=
d\tilde{Z}^M \hat{E}^{ {\tilde{a}}}_M(\tilde{Z}) \qquad 
\\ \label{KKB} & type \; IIB \; : & 
E^a = ({E}^{\tilde{a}}, {E}^{*}) \; , \qquad 
{E}^{ {\tilde{a}}}= {E}^{ {\tilde{a}}(-)} = 
d\tilde{Z}^M  {E}^{ {\tilde{a}}}_M(\tilde{Z}) \; , \qquad 
\end{eqnarray}
the T--duality rules for 
NS-NS superfields have the form: for the bosonic supervielbeins 
\begin{eqnarray}\label{Ts1}
e^{{\Phi(\tilde{Z})\over 4}} E^{ {\tilde{a}}(-)} =  
e^{{\hat{\Phi}(\tilde{Z})\over 4}}\hat{E}^{ {\tilde{a}}(-)} \; , 
\qquad 
e^{{\Phi\over 4}} E_y^{ {*}}= {1\over 
e^{{\hat{\Phi}\over 4}} 
\hat{E}_{\hat{z}}^{ {\#}}}\; , 
\qquad  
e^{{\Phi\over 4}} E^{ {*}(-)}= {i_{\hat{z}} 
\hat{B}_2 \over e^{{\hat{\Phi}\over 4}} \hat{E}_{\hat{z}}^{ {\#}}} \; , 
\end{eqnarray}
for dilaton 
\begin{eqnarray}\label{Ts1P}
e^{{\Phi}(\tilde{Z})} = {e^{\hat{\Phi}(\tilde{Z})}\over 
e^{{\hat{\Phi}\over 4}} 
\hat{E}_{\hat{z}}^{ {\#}}} \; , 
\qquad 
\end{eqnarray}
and for the NS-NS superforms
\begin{eqnarray}\label{Ts2} 
i_yB_2 = {\hat{E}^{ {\#}(-)}\over 
\hat{E}_{\hat{z}}^{ {\#}}} \; , 
\qquad && 
B_2^{(-)}= \hat{B}^{(-)}_2 - i_{\hat{z}}\hat{B}_2\wedge 
 {\hat{E}^{ {\#}(-)}\over \hat{E}_{\hat{z}}^{ {\#}}} \; . 
\end{eqnarray}

The T--duality rules for fermionic supervielbeins are
\begin{eqnarray}\label{Tsf19}
e^{-{1\over 8}{\Phi}} 
{{E}_y^{{\beta}1} \over E_y^*} 
&=& - \,  e^{-{1\over 8}\hat{\Phi}}\; 
\left(
{\hat{E}_{\hat{z}}^{ {{\b}1}}\over \hat{E}_{\hat{z}}^{ {\#}}} 
+ {i \over 4} 
\tilde{\sigma}^{ {\#}}{}^{ {\beta} {\gamma}}
\hat{\nabla}_{ {\gamma}1}\hat{\Phi}- {i \over 8} 
\tilde{\sigma}^{ {\#}}{}^{ {\beta} {\gamma}}
\hat{\nabla}_{ {\gamma}1} ln \left(e^{\hat{\Phi}\over 4} E_{\hat{z}}^{\#}
\right)\right) \; , \qquad \\  
\label{Tsf29}
e^{-{1\over 8}{\Phi}} 
{{E}_y^{ {\beta}2}\over E_y^*} &=&
\; e^{- {1\over 8}\hat{\Phi}}\;  
\tilde{\sigma}^{ {\#} {\beta} {\gamma}}
\; \left({\hat{E}_{\hat{z}}{}_{\gamma}^2 \over \hat{E}_{\hat{z}}^{ {\#}}} 
+ {i \over 4}  
{\sigma}^{\#}_{ {\beta} {\gamma}}
\hat{\nabla}^{ {\gamma}}_2\hat{\Phi} - 
{i \over 8}  
{\sigma}^{\#}_{ {\beta} {\gamma}}
\hat{\nabla}^{ {\gamma}}_2
ln \left({e^{\hat{\Phi}\over 4}} E_{\hat{z}}^{\#}\right)
\right) \; ,  \qquad \\ 
\label{Tsf1-}
e^{{1\over 8}{\Phi}} 
{E}^{ {\beta}1[-]} 
&=& \;  e^{{1\over 8}\hat{\Phi}}\; 
\left(\hat{E}^{{\beta}1[-]} - {i \over 8} \hat{E}^{ {\tilde{a}}(-)}
\tilde{\sigma}_{ {\tilde{a}}}{}^{ {\beta} {\gamma}}
\hat{\nabla}_{ {\gamma}1} ln \left({e^{\hat{\Phi}\over 4}} E_{\hat{z}}^{\#}
\right)
\right)
\; , \qquad \\  
\label{Tsf2-} 
e^{{1\over 8}{\Phi}} 
{E}^{ {\beta}2[-]}
&=& \; e^{{1\over 8}\hat{\Phi}}\;  
\tilde{\sigma}^{ {\#} {\beta} {\gamma}}
\; \left( \hat{E}_{ {\gamma}}^{2[-]}  - {i \over 8}  
\hat{E}^{{\tilde{a}}(-)}
{\sigma}_{\tilde{ {a}} {\beta} {\gamma}}
\hat{\nabla}^{ {\gamma}}_2
ln \left({e^{\hat{\Phi}\over 4}} E_{\hat{z}}^{\#}\right)\right)
\; . \qquad 
\end{eqnarray}

T--duality rules for the RR superform potentials are
\begin{eqnarray}
\label{Ts3} 
C^{(0)} &=&  i_{\hat{z}}\hat{C}_1 \equiv \hat{C}^{(1)}_{\hat{z}}\; , 
\nonumber \\ 
i_{y}C_{2n}  &=& - \hat{C}^{(-)}_{2n-1} + 
{\hat{E}^{ {\#}(-)}\over \hat{E}_{\hat{z}}^{ {\#}}}\,\wedge 
i_{\hat{z}} \hat{C}_{2n-1} \; , \quad n=1,2,3,4,5\; ,  
\nonumber \\ 
C^{(-)}_{2n} 
&=& i_{\hat{z}}\hat{C}_{2n+1} + 
i_{\hat{z}} \hat{B}_2 \wedge 
\left(\hat{C}^{(-)}_{2n-1} -  \,
{\hat{E}^{ {\#}(-)}\over \hat{E}_{\hat{z}}^{ {\#}}}\, 
\wedge 
i_{\hat{z}} \hat{C}_{2n-1} \right)\; , 
\\ \nonumber && \quad n=1,2,3,4\; .
\end{eqnarray}

\section*{A2. 
T-duality rules with supervielbeins in the string frame 
which are not adapted to the isometry} 

The relation between Einstein frame and string frame supervielbein is given in 
Eq. (\ref{Estr}). 
In the string frame the T--duality rules for  NS-NS superfields acquire 
the form (the supervielbein forms are not assumed to be adapted to the 
isometries as in Appendix A1)   
\begin{eqnarray}\label{TE-nadE}
&& {\cal E}^{a(-)}= 
\hat{{\cal E}}^{a(-)} + 
{\hat{{\cal E}}^{a(-)}\hat{{\cal E}}_{\hat{z}\, a}\over 
\hat{{\cal G}}_{\hat{z}\hat{z}} }
- {i_{\hat{z}} \hat{B}_2 \over 
\hat{{\cal G}}_{\hat{z}\hat{z}} }  {\cal E}_{\hat{z}}^a  \; , 
\qquad  i_y {\cal E}^{a}= - {\hat{{\cal E}}_{\hat{z}}^{a}
\over \hat{{\cal G}}_{\hat{z}\hat{z}}}\; , \quad \\ 
\label{TPnadE}
&& e^{2{\Phi}} = - {e^{2\hat{\Phi}(\tilde{Z})}\over 
\hat{{\cal G}}_{\hat{z}\hat{z}}}\; ,  \quad \\ 
\label{TBnadE}
&& B_2^{(-)} = \hat{B}^{(-)}_2 - i_{\hat{z}}\hat{B}_2\wedge 
 {\hat{{\cal E}}^{a(-)} \hat{{\cal E}}_{{\hat{z}}a}
\over \hat{{\cal G}}_{\hat{z}\hat{z}}} \; ,  \qquad    
 i_yB_2 = {\hat{{\cal E}}^{a(-)} \hat{{\cal E}}_{{\hat{z}}a}
\over 
\hat{{\cal G}}_{\hat{z}\hat{z}}}\; . \quad 
\end{eqnarray}
The T--duality rules for the fermionic supervielbeins are  
\begin{eqnarray}\label{Tf-nad}
{\cal E}_y^{\beta 1} &=& {\hat{{\cal E}}_{\hat{z}}^{{\beta}1}\over 
\hat{{\cal G}}_{\hat{z}\hat{z}}}\; , \qquad 
{\cal E}_y^{\beta 2} =  
{\hat{{\cal E}}_{\hat{z}}^{{a}}\tilde{\sigma}_a^{\beta\gamma}
\hat{{\cal E}}_{\hat{z}}{}_{\gamma}^{2}\over 
\sqrt{|\hat{{\cal G}}_{\hat{z}\hat{z}}|}
\hat{{\cal G}}_{\hat{z}\hat{z}}}\; , \quad 
\\ \nonumber 
 {\cal E}^{\beta 1(-)} &=& \hat{{\cal E}}^{\beta 1(-)}
- {\hat{{\cal E}}^{a(-)} \hat{{\cal E}}_{\hat{z}\, a}
\over \hat{{\cal G}}_{\hat{z}\hat{z}} }
\, \hat{{\cal E}}_{\hat{z}}^{{\beta}1} +  
{i_{\hat{z}} \hat{B}_2 \over \hat{{\cal G}}_{\hat{z}\hat{z}}}
\, \hat{{\cal E}}_{\hat{z}}^{{\beta}1}
\; , \quad \\ \nonumber 
 {\cal E}^{\beta 2(-)} &=& -  { \hat{{\cal E}}_{\hat{z}}^{\hat{a}} 
\tilde{\sigma}_a^{\beta\gamma} \over \sqrt{|\hat{{\cal G}}_{\hat{z}\hat{z}}|}}
\left(\hat{{\cal E}}_{\gamma}^{2(-)} -
{\hat{{\cal E}}^{a(-)} \hat{{\cal E}}_{\hat{z}a} 
\over \hat{{\cal G}}_{\hat{z}\hat{z}} } \, 
\hat{{\cal E}}_{\hat{z}}{}_{\gamma}^{2(-)}
- {i_{\hat{z}} \hat{B}_2 \over \hat{{\cal G}}_{\hat{z}\hat{z}}} 
\, \hat{{\cal E}}_{\hat{z}}{}_{\gamma}^{2}\right) . \quad 
\end{eqnarray}
The T--duality rules for the RR--superform potentials are 
\begin{eqnarray}
\label{TCr0} 
C^{(0)} &=&  i_{\hat{z}}\hat{C}_1 \equiv \hat{C}^{(1)}_{\hat{z}}\; , 
\\ \label{TCr2}
i_{y}C_{2n}  &=& - \hat{C}^{(-)}_{2n-1} +  
{\hat{{\cal E}}^{a(-)}\hat{{\cal E}}_{\hat{z}a}
\over \hat{{\cal G}}_{\hat{z}\hat{z}}}
\,\wedge 
i_{\hat{z}} \hat{C}_{2n-1} \; , \quad n=1,2,3,4,5\; ,
\\ \label{TCr3}
C^{(-)}_{2n} 
&=& i_{\hat{z}}\hat{C}_{2n+1} + 
i_{\hat{z}} \hat{B}_2 \wedge 
\left(\hat{C}^{(-)}_{2n-1} - \, {\hat{{\cal E}}^{a(-)}\hat{{\cal E}}_{\hat{z}a}
\over \hat{{\cal G}}_{\hat{z}\hat{z}}} \,
\wedge 
i_{\hat{z}} \hat{C}_{2n-1} \right)\, ,  \\ 
&& \nonumber \quad n=1,2,3,4\; .
\end{eqnarray} 
Here
\begin{eqnarray}\label{calG-r}
\hat{{\cal G}}_{\hat{z}\hat{z}}\equiv \hat{{\cal E}}_{\hat{z}}^{\hat{a}}
\hat{{\cal E}}_{\hat{z}\hat{a}}\; .
\end{eqnarray}

\newpage 

\section*{A3. The complete 
differential form representation for the T--duality rules 
(\ref{Ts2})--(\ref{Ts3})} 

For NS--NS superforms: 
\begin{equation}\label{TBsg-r1}
{{B}}_2=  \hat{{B}}_2 -
(dy + i_{\hat{z}} \hat{B}_2) \wedge
\left(d\hat{z}+ {\hat{E}^{ {\#}(-)}
\over \hat{E}_{\hat{z}}^{ {\#}}}\right) + dy \wedge d\hat{z}\; ,    
\end{equation} 
or  
\begin{equation}
\label{rTdB2-r2} 
B_2 = \hat{B}_2 - \, 
{1\over {E}_y^{ {*}}} \, 
E^{ {*}} \wedge \hat{E}^{ {\# }} \,
{1\over \hat{E}_{\hat{z}}^{ {\#}}}\, 
+ dy  \wedge d\hat{z} \; .
\end{equation}

For RR--superforms: 
\begin{equation}
\label{rC=-r1} 
C= i_{\hat{z}}\hat{C} + (dy +i_{\hat{z}} \hat{B}_2) \wedge 
\left(\hat{C}^{(-)} -  \,
{\hat{E}^{ {\#}(-)}\over \hat{E}_{\hat{z}}^{ {\#}}}\, 
\wedge 
i_{\hat{z}} \hat{C} \right)\, ,
\end{equation}
or 
\begin{equation}\label{rhC=r1}
\hat{C} = - i_y C + (d\hat{z}+ i_yB_2)\wedge 
\left(C^{(-)}- i_y C \wedge \,
 {{E}^{ {*}(-)}\over {E}_y^{ {*}}}\, \right)
\, .  
\end{equation}

For fermionic supervielbeins: 
\begin{eqnarray}\label{Td-EhE1r!}
e^{{1\over 8}{\Phi}} 
({E}^{ {\beta}1} - {i \over 8} 
{E}^{ {a}} 
\tilde{\sigma}_{ {a}}{}^{ {\beta} {\gamma}}
{\nabla}_{ {\gamma}1}{\Phi}) 
= \;  e^{{1\over 8}\hat{\Phi}}\; 
(\hat{E}^{ {\beta}1} - {i \over 8} 
\hat{E}^{ {a}}
\tilde{\sigma}_{ {a}}{}^{ {\beta} {\gamma}}
\hat{\nabla}_{ {\gamma}1}\hat{\Phi}) - 
\;  \qquad \nonumber \\  
{} \qquad - 
 \;  e^{{1\over 8}\hat{\Phi}}\; 
\left(\hat{E}^{ {\#}} + 
e^{{1\over 4}(\Phi -\hat{\Phi})}
{E}^{ {*}}\right) \; \left(
{
\hat{E}_{\hat{z}}^{ {\beta}1}\over \hat{E}_{\hat{z}}^{ {\#}}}
+  {i \over 8} 
\tilde{\sigma}^{ {\#}}{}^{ {\beta} {\gamma}}
\hat{\nabla}_{ {\gamma}1}\hat{\Phi}\right) \; , 
\\ 
\label{Td-EhE2r!} 
e^{{1\over 8}{\Phi}} 
({E}^{ {\beta}2} - {i \over 8} 
{E}^{ {a}} 
\tilde{\sigma}_{ {a}}{}^{ {\beta} {\gamma}}
{\nabla}_{ {\gamma}2}{\Phi}) 
= \; 
e^{{1\over 8}\hat{\Phi}}\;  
\tilde{\sigma}^{ {\#} {\beta} {\gamma}}
\; ( \hat{E}_{ {\gamma}}^2  - {i \over 8}  
\hat{E}^{ {a}}
{\sigma}_{{ {a}} {\beta} {\gamma}}
\hat{\nabla}^{ {\gamma}}_2\hat{\Phi}) - 
 \qquad \nonumber \\  
{} \qquad -   e^{{1\over 8}\hat{\Phi}}\,  
\left(\hat{E}^{ {\#}} - 
e^{{1\over 4}(\Phi -\hat{\Phi})}
{E}^{ {*}}\right) \, 
\tilde{\sigma}^{ {\#} {\beta} {\gamma}}
\left({\hat{E}_{\hat{z}}{}_{ {\gamma}}^2
\over \hat{E}_{\hat{z}}^{ {\#}}}  + {i \over 8}  
{\sigma}^{\#}_{ {\beta} {\gamma}}
\hat{\nabla}^{ {\gamma}}_2\hat{\Phi}\right) \; . 
\end{eqnarray}

\section*{A4. T-duality for the field strengths} 

For NS--NS superfield strengths $\hat{H}_3= d \hat{B}_2$ 
and $H_3=dB_2$ the T--duality rules read
\begin{equation}
\label{Td-HH1} 
H_3 = \hat{H}_3 
-  \left(d\hat{z} + 
{\hat{E}^{ {\#}(-)}\over \hat{E}_{\hat{z}}^{ {\#}}} 
\right)\wedge i_{\hat{z}} \hat{H}_3 \; + 
 \left(dy +  {{E}^{ {*}(-)}\over {E}_y^{ {*}}} \,
\right)\wedge i_y H_3  \; , 
\end{equation}
or, equivalently,   
\begin{eqnarray}
\label{Td-HH2} &
H_3 = & \hat{H}_3 
- {1\over {E}_y^{ {*}}\, \hat{E}_{\hat{z}}^{ {\#}}}
\, E^{ {*}} \wedge \hat{T}^{ {\# }}  
+ {1\over {E}_y^{ {*}}\, \hat{E}_{\hat{z}}^{ {\#}}}
\, \hat{E}^{ {\#}} \wedge {T}^{ {*}} + \\ 
\nonumber 
 & & + 
{1\over {E}_y^{ {*}}\, \hat{E}_{\hat{z}}^{ {\#}}}
\left( \hat{E}^{ {\#}} \wedge 
{\cal E}^{ {\tilde{b}}}\wedge 
{w}_{ {\tilde{b}}}{}^{ {*}}
-  {E}^{ {*}} \wedge 
{E}^{ {\tilde{b}}}\wedge \hat{w}_{ {\tilde{b}}}
{}^{ {\#}} +  {E}^{ {*}} \wedge 
\hat{E}^{ {\#}}\wedge d\, 
log|{E}_y^{ {*}}\, \hat{E}_{\hat{z}}^{ {\#}}|\right) 
\; .
\end{eqnarray}

For RR superfield strengths the T--duality rules can be collected in 
the relation    
\begin{eqnarray}
  \label{Td-RhR} 
R =  -  i_{\hat{z}} \hat{R}   +  (dy +  
{E^{ {*}(-)}\over E_y^{ {*}}})\wedge 
(\hat{R} + (d\hat{z} + 
{\hat{E}^{ {\#}(-)}\over \hat{E}_{\hat{z}}^{ {\#}}})
\wedge i_{\hat{z}} \hat{R} ) \; . 
\end{eqnarray}

\end{document}